\documentclass[sn-mathphys-num]{sn-jnl}
\usepackage{graphicx}%
\usepackage{multirow}%
\usepackage{amsmath,amssymb,amsfonts}%
\usepackage{amsthm}%
\usepackage{mathrsfs}%
\usepackage[title]{appendix}%
\usepackage{xcolor}%
\usepackage{textcomp}%
\usepackage{manyfoot}%
\usepackage{booktabs}%
\usepackage{algorithm}%
\usepackage{algorithmicx}%
\usepackage{algpseudocode}%
\usepackage{listings}%

\theoremstyle{thmstyleone}%
\theoremstyle{thmstyletwo}%
\theoremstyle{thmstylethree}%
\begin{document}

\title[Article Title]{Viscoelastic wetting transition: beyond lubrication theory}

\author[1]{\fnm{Minkush} \sur{Kansal}}

\author[1,2]{\fnm{Charu} \sur{Datt}}

\author[1,3]{\fnm{Vincent} \sur{Bertin}}

\author*[1]{\fnm{Jacco H.} \sur{Snoeijer}}\email{j.h.snoeijer@utwente.nl}

\affil[1]{\orgdiv{Physics of Fluids Group, Faculty of Science and Technology}, \orgname{University of Twente}, \orgaddress{\street{P.O. Box 217}, \city{Enschede} \postcode{7500 AE}, \country{The Netherlands}}}

\affil[2]{\orgdiv{Present address: Department of Mechanical Engineering}, \orgname{Keio University}, \orgaddress{\city{Yokohama} \postcode{223-8522}, \country{Japan}}}

\affil[3]{\orgdiv{Present address: Aix Marseille Univ}, \orgname{CNRS}, \orgaddress{\street{IUSTI UMR 7343}, \city{Marseille} \postcode{13453}, \country{France}}}

\abstract{The dip-coating geometry, where a solid plate is withdrawn from or plunged into a liquid pool, offers a prototypical example of wetting flows involving contact-line motion. Such flows are commonly studied using the lubrication approximation approach which is intrinsically limited to small interface slopes and thus small contact angles. Flows for arbitrary contact angles, however, can be studied using a generalized lubrication theory that builds upon viscous corner flow solutions. Here we derive this generalized lubrication theory for viscoelastic liquids that exhibit normal stress effects and are modelled using the second-order fluid model. We apply our theory to advancing and receding contact lines in the dip-coating geometry, highlighting the influence of viscoelastic normal stresses for contact line motion at arbitrary contact angle.}

\maketitle

\section{Introduction}
\label{sec:introduction}

The dip-coating process consists of dragging an object out of a liquid bath and is widely used to deposit a uniform thin coating onto a surface in industry \citep{scriven1988physics, schyrr2014biosensors, riau2016functionalization}. An important body of research has been devoted to the prediction of the deposited film thickness since the seminal work of Landau and Levich~\cite{levich1942dragging,quere1999fluid,rio2017withdrawing,bertin2022enhanced}. For non-wetting liquids, the entrainment of a thin liquid film occurs only if the plate speed $U$ exceeds a critical speed $U_c$~\cite{gennes2004capillarity}. On the other hand for $U<U_c$, which will be the regime of interest in this article, the meniscus is shifted as compared to the equilibrium stationary shape leading to displacement of the contact-line \cite{eggers2005existence,snoeijer2007relaxation,chan2012theory,galvagno2014continuous,gupta2023experimental}. Hence, the flows in the dip-coating geometry are a prototypical example of the contact-line flows that have interested the soft matter community for decades because the wetting transitions they display are also of relevance to other processes like inkjet printing and spray coating~\cite{de1985wetting,bonn2009wetting,snoeijer2013moving}. Many of these applications utilize non-Newtonian fluids that exhibit properties like shear thinning viscosity and viscoelasticity. In this article, we focus on viscoelastic liquids such as polymeric solutions that show significant normal-stress differences~\cite{bird1987dynamics}. The viscoelastic normal stresses are expected to be significant for the contact line motion due to the high shear rates~\cite{bird1987dynamics,tanner2000engineering,bartolo2007dynamics}, as also demonstrated in our recent work~\citep{datt2022thin,kansal2024viscoelastic}. However, moving contact line experiments with polymeric liquids~\cite{seevaratnam2007dynamic,wei2009dynamic,smith2014origin,shin2015contact} and numerical simulations using dumbbell rheological models~\cite{yue2012phase,wang2015dynamic} have revealed a complex dynamics that may also depend on shear-thinning effects, polymer concentration, and polymer-surface interactions. One difficulty to interpret those dynamics is to disentangle the various polymeric effects, notably because of the lack of established theoretical framework for viscoelastic contact line-motion. 

We recently derived theoretically a modified Cox-Voinov theory of moving contact line that includes viscoelastic normal stress differences~\cite{kansal2024viscoelastic}, by performing asymptotic analysis of the viscoelastic thin-film model of~\cite{datt2022thin}. The latter was derived using the second-order fluid model, which is the simplest frame-invariant rheological model to study viscoelastic fluids exhibiting normal stress differences. We note that the second-order fluid provides a good description of slow and steady flows as is typically the case for moving contact lines, though it often leads to unphysical results in strongly unsteady or high-Weissenberg flows~\citep{tanner2000engineering,morozov2015introduction,decorato16b}. Nevertheless, the one limitation of the viscoelastic Cox-Voinov expression of \cite{kansal2024viscoelastic} is that it is \textit{a priori} restricted to small contact angles, as it is derived from a thin-film model and it relies on the small slope approximation. The aim of this article is to generalize the viscoelastic contact-line model to large angles, as are encountered in systems like in dip-coating. 

\begin{figure}
	\centerline{\includegraphics[width=\linewidth]{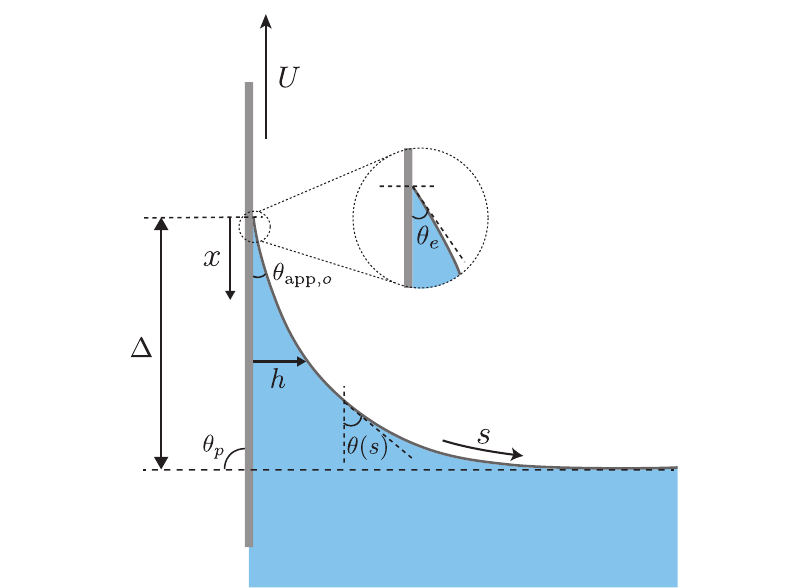}}
	\caption{Dip-coating geometry. The interface can be described by $h(s)$ and $\theta(s)$. The plate is being pulled at an angle $\theta_p$ (here $\pi/2$) with respect to the stationary fluid bath. The interface depth from the contact line to the bath is denoted by $\Delta$, which on the scale of the meniscus has an apparent (outer) angle $\theta_{{\rm app},o}$. The microscopic angle at the contact line is $\theta_e$, as shown in the zoomed picture.}
	\label{fig:fig1}
\end{figure}	

The problem studied in this article is depicted in figure~\ref{fig:fig1}. A plate is being pulled out of a liquid bath at a steady speed $U$ with an angle $\theta_p$. The liquid wets the plate with a microscropic contact angle $\theta_e$. We assume $U < U_c$ such that the meniscus has a stationary solution parametrized by a film thickness $h(s)$ and local slope angle $\theta(s)$, where $s$ is the curvilinear coordiate. To obtain the meniscus shape and ultimately the contact line motion (quantified by the meniscus height $\Delta$) at large contact angles, we use the generalized lubrication theory~\cite{snoeijer2006free}, which is based on the corner flow solutions of viscous flows~\citep{huh1971hydrodynamic,cox_1986}. The central assumption behind this analysis is that the interface angle $\theta(s)$ is slowly varying; this requires a small curvature of the interface, but no restriction on the magnitude of $\theta(s)$. As done for Newtonian fluids in~\citep{snoeijer2006free}, similar corner flows can be constructed for second-order fluids~\cite{han2014theoretical}, which offers a route to develop a generalized lubrication theory that includes normal stresses.  

The paper is structured as follows. We first compute the corner flow solutions for the second-order fluid model in section~\ref{sec:corner_flow}, before formulating the generalized lubrication theory with viscoelastic normal stresses in section~\ref{sec:generalized_lubrication_equation}. In section~\ref{sec:application_to_dip_coating}, we apply the generalized lubrication theory to dip-coating for both receding (pulling the plate) and advancing (pushing the plate) contact lines. Some concluding remarks are given in section~\ref{sec:conclusion}.

\section{Corner flow in a second-order fluid}
\label{sec:corner_flow}

\begin{figure}
	\centerline{\includegraphics[width=\linewidth]{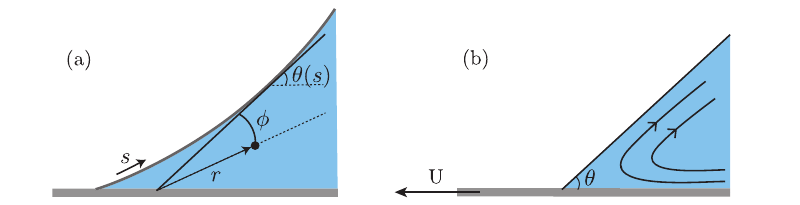}}
	\caption{Corner flow geometry: (a) Definition of the polar coordinates $r$ and $\phi$ in the locally tangent wedge of angle $\theta(s)$. At this tangent position $s$ along the interface, we use the corner flow solution (see panel (b)). Here, the interface point corresponds to $\phi=0$ and the plate is at $\phi=\theta(s)$. (b) Illustration of the streamlines in the corner flow solution, in a wedge of constant angle $\theta=\theta(s)$, sketched in the frame comoving with the interface.}
	\label{fig:fig2_schematic}
\end{figure}

The generalized lubrication approach is based on viscous corner flows, for which exact solutions have been available since the work by \citet{huh1971hydrodynamic}. More specifically, at a given location $s$ along the interface (see figure~\ref{fig:fig2_schematic}(a)), we approximate the flow with the corner flow solution at an angle $\theta(s)$, where $\theta$ is the local interface angle. In this section we first recall the viscous corner flow solutions and their properties in subsection~\ref{sec:huh_scriven}, and subsequently we extend the solution by adding the normal-stress effect using the second-order model as a fluid constitutive relation in subsection~\ref{sec:second_order_fluid}. 

The corner flow problem is sketched in figure~\ref{fig:fig2_schematic}(b). For both Newtonian and viscoelastic fluids, the flow is assumed incompressible and two-dimensional, so we introduce the stream-function $\psi (r, \phi)$, which gives the radial and azimuthal velocity components
\begin{equation}
	\label{eq:stream_function-velocity-definition}
	u_r= \frac{1}{r} \frac{\partial \psi}{\partial \phi}, \quad \mathrm{and} \quad u_\phi = - \frac{\partial \psi}{\partial r}.
\end{equation}
The boundary conditions for this steady corner flow are that the domain boundaries are streamlines, i.e. no fluid penetration through the boundaries, and a no-slip boundary condition is imposed at the moving boundary at $\phi = \theta$. This leads to velocity boundary conditions

\begin{equation}\label{eq:vbc}
	\left. u_\phi \right|_{\phi=0}= 0, \quad \left. u_\phi \right|_{\phi=\theta}= 0, \quad \left. u_r \right|_{\phi=\theta}= -U, 
\end{equation}
where $U$ is the plate speed in the frame of the steady corner flow. The sign convention here is that $U$ is taken positive for a receding contact line (cf. figure~\ref{fig:fig2_schematic}). The free surface is assumed to be subjected to a no-shear boundary condition, 
\begin{equation}\label{eq:sbc}
	\left. \sigma_{\phi r} \right|_{\phi=0}= 0,
\end{equation}
which is expressed in terms of the off-diagonal component of the stress tensor $\pmb \sigma$. Differences between the Newtonian and the second-order fluid arise in the constitutive relation, relating stress to the rate-of-strain tensor

\begin{equation}
	\dot{\pmb{\gamma}} = \nabla \pmb{u} + \left(\nabla \pmb{u} \right)^T ,
\end{equation}
as will be specified below. These differences manifest themselves both in the bulk (through momentum equation, which upon neglecting inertia and body forces becomes $\nabla \cdot \pmb \sigma = \mathbf 0$), and at the boundary (specifically via~\eqref{eq:sbc}). 

\subsection{Newtonian fluid: Huh-Scriven solutions} 
\label{sec:huh_scriven}

For Newtonian flow the fluid stress is given by $\pmb{\sigma}= -p \mathrm{\textbf{I}}+ \eta \dot{\pmb{\gamma}}$, and for a flow without inertia the momentum equation reduces to the Stokes equation, 

\begin{equation}\label{eq:stokes}
	- \nabla p+ \eta \nabla^2 \pmb{u} =0.
\end{equation}
Taking the curl $\left(\nabla\times\right)$ of the Stokes equation eliminates the pressure and gives the biharmonic equation for $\psi$, 
\begin{equation}
	\nabla^4 \psi = 0. 
\end{equation}
This fourth-order equation requires four boundary conditions, which are provided by \eqref{eq:vbc} and \eqref{eq:sbc}. In terms of the stream-function, the velocity boundary conditions \eqref{eq:vbc} can be expressed as
\begin{eqnarray} \label{eq:vbc_psi_form}
	\left. \frac{\partial \psi}{\partial r} \right|_{\phi=0}= 0, \quad \left. \frac{\partial \psi}{\partial r} \right|_{\phi=\theta}= 0, \quad\left. \frac{1}{r}\frac{\partial \psi}{\partial \phi} \right|_{\phi=\theta}= - U. 
\end{eqnarray}
The vanishing shear stress boundary condition \eqref{eq:sbc} for a Newtonian fluid reduces to,

\begin{equation} \label{eq:sbc_psi_form}
	\left. \frac{1}{r}\frac{\partial^2 \psi}{\partial \phi^2} \right|_{\phi=0} =0.
\end{equation}

The problem defined above has been solved exactly~\citep{huh1971hydrodynamic}, using a similarity function of the type $\psi(r,\phi)= r^\alpha F(\phi)$. The inhomogenous no-slip velocity boundary condition implies $\alpha=1$, which determines the functional form of $F(\phi)$. Using the boundary conditions~\eqref{eq:vbc_psi_form} and~\eqref{eq:sbc_psi_form}, one then finds 

\begin{equation}\label{eq:HSpsi}
	\psi (r,\phi) = \frac{U \sin{\theta}}{\theta - \cos{\theta} \sin{\theta}} r \phi \cos{\phi} - \frac{U \theta \cos{\theta}}{\theta - \cos{\theta} \sin{\theta}} r \sin{\phi}.
\end{equation}
From this stream function, the velocity components can be expressed by using \eqref{eq:stream_function-velocity-definition} and gives
\begin{equation}
	\label{eqn:velocity_u_r}
	u_r= U \frac{(\cos{\phi}- \phi \sin{\phi}) \sin{\theta} - \theta \cos{\theta} \cos{\phi}}{\theta - \cos{\theta} \sin{\theta}},
\end{equation}
\begin{equation}
	\label{eqn:velocity_u_phi}
	u_\phi= U \frac{\theta \sin{\phi} \cos{\theta}- \phi \cos{\phi} \sin{\theta}}{\theta - \cos{\theta} \sin{\theta}}. 
\end{equation}

The central observation made by \citet{huh1971hydrodynamic} is that the velocity field in this corner flow gives rise to a singularity of the stress $\sim \eta U/r$, which has dramatic consequences for moving contact lines. This scaling could have been anticipated via dimensional analysis, given that the flow structure is self-similar and lacks any length-scale apart from $r$. An explicit expression for the stress tensor involves the pressure inside the fluid, which can be obtained by integration of \eqref{eq:stokes}. As it will be useful later in the paper, when capillarity is introduced, we write here the resulting pressure at the free surface ($\phi=0$), 

\begin{equation}\label{eqn:pN}
	\left. p \right|_{\phi=0}= \frac{2 \eta U}{r} \left(\frac{\sin{\theta}}{\theta-\cos{\theta}\sin{\theta}}\right) + p_\infty,
\end{equation}
where $p_\infty$ is the reference pressure at infinity. One verifies from the velocity solution that $\dot \gamma_{\phi \phi}=0$ at $\phi=0$, such that the normal stress at the free surface reads

\begin{equation}
	\label{eqn:normal_stress_free_surface_Newtonian}
	\left( \mathbf n \cdot \pmb \sigma \cdot \mathbf n \right)_{\phi=0}= - p,
\end{equation} 
where $\mathbf n$ is the normal vector to the interface. This result points to an important subtlety of wetting flows: with Laplace's law, such a normal stress (that is a function of $r$, as given by \eqref{eqn:pN}) is expected to lead to a curved interface~\citep{huh1971hydrodynamic}. Yet, the interface was assumed to be perfectly straight. Therefore, the corner flow can only be viewed as an approximation in the limit of weakly curved interfaces, which amounts to demanding the capillary number to be small.

\subsection{Second-order fluid} 
\label{sec:second_order_fluid}

\subsubsection{Velocity field}

We extend our analysis of corner flows to viscoelastic fluids using the second-order fluid model, as discussed in section~\ref{sec:introduction}. Writing the stress as $\pmb \sigma = - p \mathbf I + \pmb \tau$, the deviatoric stress $\pmb \tau$ of the second-order fluid is given by \citep{tanner2000engineering,morozov2015introduction}
\begin{equation}
	\label{eqn:tau_second_order_fluid}
	\pmb{\tau}= \eta \dot{\pmb{\gamma}} - \frac{\psi_1}{2} \buildrel \nabla \over {\dot{\pmb{\gamma}}} + \psi_2 \dot{\pmb{\gamma}} \cdot \dot{\pmb{\gamma}}, 
\end{equation}
where we introduced the upper-convected derivative
\begin{equation}
	\label{eqn:upper_convected_term}
	\buildrel \nabla \over {\dot{\pmb{\gamma}}}=  \frac{\partial \dot{\pmb{\gamma}}}{\partial t} + \pmb{u} \cdot \nabla \dot{\pmb{\gamma}} - \left(\nabla \pmb{u} \right)^T \cdot \dot{\pmb{\gamma}} - \dot{\pmb{\gamma}} \cdot \nabla \pmb{u}. 
\end{equation}
Besides the usual viscous contribution, the constitutive relation \eqref{eqn:tau_second_order_fluid} contains extra terms that represent normal stress differences, $\psi_{1,2}$ representing the first and second normal stress coefficients respectively. 

To analyse the flow, one can once again evaluate the momentum equation $\nabla \cdot \pmb \sigma = \mathbf 0$ and eliminate the pressure by taking the curl. For steady flows, the seminal work by \citet{tanner1966plane} demonstrated that the resulting equation for the stream-function reads
\begin{equation}
	\eta \nabla^4 \psi - \frac{\psi_1}{2} \pmb{u} \cdot \nabla \left(\nabla^4 \psi \right) =0.
\end{equation}
This equation has a remarkable consequence: any viscous flow solution, which has the property that $\nabla^4 \psi = 0$, automatically satisfies the momentum equation of a second-order fluid. This observation has led to a theorem, articulated by \citet{tanner1966plane} as: \emph{``Any plane creeping Newtonian velocity field with given velocity boundary conditions is also a solution for the second-order incompressible fluid with the same boundary conditions."} 

\citet{han2014theoretical} considered that the Newtonian corner solution by \citet{huh1971hydrodynamic} is also applicable to the second-order fluid, owing to Tanner's theorem. While this assertion turns out to be correct, the applicability of  \eqref{eq:HSpsi} is not clear \emph{a priori} as the problem involves not only velocity boundary conditions \eqref{eq:vbc} but also a stress boundary condition \eqref{eq:sbc} and the stress contains extra, viscoelastic terms. The velocity boundary conditions are naturally respected by \eqref{eq:HSpsi}, but we need to verify explicitly whether the solution satisfies $\sigma_{\phi r}=\tau_{\phi r}=0$ at the interface. In Appendix~\ref{sec:appendix}, we show that the shear stress with the second order fluid model reduces to
\begin{align} 
	\label{eq:shear-stress_second-order}
	\tau_{\phi r}  =& \dfrac{\eta}{r} \left(\dfrac{\partial u_r}{\partial \phi} - u_\phi \right) - \frac{\psi_1}{2} \bigg[ -\dfrac{u_r}{r^2} \left(\dfrac{\partial u_r}{\partial \phi} + u_\phi \right) + \dfrac{u_\phi}{r^2} \left(\dfrac{\partial^2 u_r}{\partial \phi^2} - 3 \dfrac{\partial u_\phi}{\partial \phi} \right) \notag \\
	&- \dfrac{3}{r^2} \left(\dfrac{\partial u_\phi}{\partial \phi} + u_r \right) \left(\dfrac{\partial u_r}{\partial \phi} - u_\phi\right) \bigg] + \psi_2 \left[ \dfrac{2}{r^2} \left(\dfrac{\partial u_r}{\partial \phi} - u_\phi \right) \left( \dfrac{\partial u_\phi}{\partial \phi} + u_r \right) \right] .	
\end{align}
We recall that $u_\phi=0$ at the free surface ($\phi=0$). The first term of \eqref{eq:shear-stress_second-order} is identical to that of the viscous stress, which for the no-shear boundary condition in the Huh-Scriven solution implies $\partial u_r/\partial \phi=0$ at the free surface. From this, one verifies that the terms involving $\psi_1$ and $\psi_2$ both vanish as well at the free surface, so that
\begin{equation}
	\left. \tau_{\phi r} \right|_{\phi=0} = 0.
\end{equation}
Hence, we conclude that \eqref{eq:HSpsi} is indeed the solution to the corner flow problem considered here: it satisfies both the momentum equation and all boundary conditions for the second-order fluid. 

\subsubsection{Pressure and pressure gradient}
As we will discuss in the next section~\ref{sec:full_equation}, we require $\partial p/\partial r$ in order to obtain the generalised lubrication equation. From $\partial p/\partial r$, we can once again evaluate the pressure at the free surface for the corner flow, as obtained above by integration of the Newtonian momentum equation. The pressure was previously determined by \citet{han2014theoretical}, using directly the modified pressure expression of \citet{tanner1989some}: 
\begin{equation}
	p = p_N - \frac{\psi_1}{2\eta}\frac{\mathrm{D}p_N}{\mathrm{D}t}+ \left(\frac{\psi_1}{8}+\frac{\psi_2}{2} \right) \pmb{\mathrm{tr}} \left(\dot{\pmb{\gamma}}\cdot \dot{\pmb{\gamma}} \right),
	\label{eq:Tannner_Thm}
\end{equation}
where $p_N$ is the pressure of the Newtonian problem satisfying $\nabla p_N = \eta \nabla^2\pmb{u}$ with the same velocity boundary conditions. We here confirm the resulting pressure in \citet{han2014theoretical} by explicit integration of the momentum equation, which for the second-order fluid reads
\begin{equation}
	\label{eqn:cauchy_eqn_second_order_fluid}
	- \nabla p + \eta \nabla^2 \pmb{u} + \nabla \cdot \left(- \frac{\psi_1}{2} \buildrel \nabla \over {\dot{\pmb{\gamma}}} + \psi_2 \dot{\pmb{\gamma}} \cdot \dot{\pmb{\gamma}} \right) =0. 
\end{equation}
We proceed by evaluating the radial components of \eqref{eqn:cauchy_eqn_second_order_fluid}. The radial component of $\nabla \cdot \buildrel \nabla \over {\dot{\pmb{\gamma}}}$  can be written as (see \citep{happel1983low}, page 494, equation A-9.28)
\begin{equation}
	\left( \nabla \cdot \buildrel \nabla \over {\dot{\pmb{\gamma}}} \right)_r = \frac{1}{r} \frac{\partial}{\partial r} \left( r \buildrel \nabla \over {\dot{\gamma}}_{rr} \right) + \frac{1}{r} \frac{\partial}{\partial \phi} \left( \buildrel \nabla \over {\dot{\gamma}}_{\phi r} \right) - \frac{\buildrel \nabla \over {\dot{\gamma}}_{\phi \phi}}{r}.
\end{equation}
Using the expression of $\buildrel \nabla \over {\dot{\pmb{\gamma}}}$ in polar coordinates (provided in Appendix~\ref{sec:appendix}), we find
\begin{equation}
	\begin{split}
		\left( \nabla \cdot \buildrel \nabla \over {\dot{\pmb{\gamma}}} \right)_r = &-2 u_\phi \frac{\partial u_r}{\partial \phi} + 2 u_\phi^2 - 2 \left( \frac{\partial u_r}{\partial \phi}\right)^2 - 4 u_r \frac{\partial^2 u_r}{\partial \phi^2} + 12 u_r \frac{\partial u_\phi}{\partial \phi} -2 \frac{\partial u_\phi}{\partial \phi} \frac{\partial^2 u_r}{\partial \phi^2} \\ 
		&+ 4 \left( \frac{\partial u_\phi}{\partial \phi} \right)^2 + u_\phi \frac{\partial^3 u_r}{\partial \phi^3} - 2 u_\phi \frac{\partial^2 u_\phi}{\partial \phi^2} - 3 \frac{\partial u_r}{\partial \phi} \frac{\partial^2 u_\phi}{\partial \phi^2} + 6 u_r^2.
	\end{split}
\end{equation}
Inserting the corner flow velocity field \eqref{eqn:velocity_u_r} and \eqref{eqn:velocity_u_phi}, we obtain at the interface 
\begin{equation}
	\left.  \left( \nabla \cdot \buildrel \nabla \over {\dot{\pmb{\gamma}}} \right)_r \right|_{\phi=0} =  \frac{4 U^2 \sin{\theta} \left( \sin{\theta} - \theta \cos{\theta} \right)}{r^3 \left( \theta - \cos{\theta} \sin{\theta} \right)^2}. \label{eqn:b2_method_2}
\end{equation}
In the momentum equation \eqref{eqn:cauchy_eqn_second_order_fluid}, this gives a contribution proportional to the first normal stress coefficient $\psi_1$. Likewise, for the term involving the second normal stress coefficient $\psi_2$, we need to express the radial component of $\nabla \cdot \left( \dot{\pmb{\gamma}} \cdot \dot{\pmb{\gamma}} \right)$, that is given by (see Appendix~\ref{sec:appendix}) 
\begin{align}
	\begin{split}
		\left( \nabla \cdot \left( \dot{\pmb{\gamma}} \cdot \dot{\pmb{\gamma}} \right) \right)_r =  &\frac{1}{r} \frac{\partial}{\partial r} \left( \dfrac{1}{r} \left( \dfrac{\partial u_r}{\partial \phi} - u_\phi \right)^2 \right) + \frac{1}{r} \frac{\partial}{\partial \phi} \left( \dfrac{2}{r^2} \left(\dfrac{\partial u_r}{\partial \phi} - u_\phi \right) \left( \dfrac{\partial u_\phi}{\partial \phi} + u_r \right) \right) \\  
		&- \frac{1}{r} \left( \left( \dfrac{1}{r} \left(\dfrac{\partial u_r}{\partial \phi} - u_\phi \right) \right)^2 + \left( \dfrac{2}{r} \left(\dfrac{\partial u_\phi}{\partial \phi} + u_r \right) \right)^2 \right).
	\end{split}
\end{align}
At the interface, we find 
\begin{equation}
	\left. \left( \nabla \cdot \left( \dot{\pmb{\gamma}} \cdot \dot{\pmb{\gamma}} \right) \right)_r \right|_{\phi=0} =0.  
\end{equation}
Hence, the term involving $\psi_2$ does not contribute to the pressure at the free surface. Collecting the viscous and normal stress contributions in \eqref{eqn:cauchy_eqn_second_order_fluid}, we obtain the pressure gradient along the interface, 
\begin{equation}\label{eq:dpdr}
	\left. \partial_r p \right|_{\phi=0}= -\frac{2 \eta U}{r^2} \left(\frac{\sin{\theta}}{\theta-\cos{\theta}\sin{\theta}}\right) - \frac{2 \psi_1 U^2}{r^3} \left(\frac{\sin{\theta} (\sin{\theta} - \theta \cos{\theta})}{\left(\theta-\cos{\theta}\sin{\theta}\right)^2}\right).
\end{equation}
The second term on the right hand side is the additional term due to viscoelasticity, compared to the Newtonian case. This expression \eqref{eq:dpdr} will play a central role in the development of the generalized lubrication theory developed below. 

To obtain the pressure along the interface for the corner flow, we integrate \eqref{eq:dpdr} to
\begin{equation}
	\left. p \right|_{\phi=0}= \frac{2 \eta U}{r} \left(\frac{\sin{\theta}}{\theta-\cos{\theta}\sin{\theta}}\right) + \frac{\psi_1 U^2}{r^2} \left(\frac{\sin{\theta} (\sin{\theta} - \theta \cos{\theta})}{\left(\theta-\cos{\theta}\sin{\theta}\right)^2}\right) + p_\infty.
	\label{eqn:pressure_second_order_fluid}
\end{equation}
One thus observes that the pressure is modified as compared to the Newtonian case, with an extra term $\sim \psi_1 U^2/r^2$. We further verify that $\tau_{\phi \phi}=0$ at the interface (cf. Appendix~\ref{sec:appendix}). This implies that the modified pressure directly provides the normal stress exerted onto the interface. The expression for the pressure is identical to that obtained by \citet{han2014theoretical}. We will further discuss the subtlety in approach of using pressure $p$ versus using pressure gradient $\partial p/\partial r$, in the next section. 

\section{Generalized lubrication equation} 
\label{sec:generalized_lubrication_equation}

\subsection{Derivation} 
\label{sec:full_equation}

We now move to the situation described in Fig.~\ref{fig:fig2_schematic}(a) where the interface is curved, with a varying tangent angle $\theta(s)$. The exact corner flow solutions presented above do not satisfy the normal stress boundary condition (Young-Laplace equation), 

\begin{equation}\label{eq:laplace}
	\left[ \mathbf n \cdot \pmb \sigma \cdot \mathbf n \right]_{\rm interface} = \gamma \kappa = \gamma \frac{d\theta}{ds},
\end{equation} 
which states that the jump of pressure across the interface leads to a curvature $\kappa$ of the interface. Namely, the pressure obtained for corner flows is inhomogeneous and thus must lead to variations in curvature; yet, the corner flow assumed the free surface to be completely flat ($\kappa=0$). Having said that, in the limit of very strong surface tension (${\rm Ca} \ll 1$)~\citep{snoeijer2006free}, the interface curvature will remain small and one can perform a perturbation expansion around the corner flow. 

In the spirit of a perturbation expansion, \citet{han2014theoretical} proposed to compute the interface shape using \eqref{eq:laplace}, while estimating the liquid pressure directly from \eqref{eqn:pressure_second_order_fluid} using the local interface slope $\theta(s)$. In general, however, a local expansion cannot be based on the pressure: the pressure is a \emph{non-local} quantity that is obtained by integration of $\nabla p$ over the entire domain. Indeed, \eqref{eqn:pressure_second_order_fluid} was obtained by integrating $\nabla p$ under the assumption that the corner angle $\theta$ is globally constant, which is a restrictive assumption given that we are interested in computing the variations of $\theta(s)$. A peculiar feature of the differential equation for the interface shape in \citet{han2014theoretical} is that it is second-order in nature, even in the Newtonian case, which is in contrast with the classical lubrication theory for moving contact lines, which amounts to a third-order differential equation. 

Here we follow the perturbation approach on the momentum equation, which involves the gradient of pressure along the interface, as outlined in~\citet{snoeijer2006free} and~\citet{chan2013hydrodynamics} for Newtonian flows. The gradient of pressure is a \emph{local} quantity, and we thus require the angle $\theta$ to be approximately constant only locally, not globally. For that reason, we rather base the analysis on the derivative of \eqref{eq:laplace} along the curvilinear coordinate $s$, i.e. 

\begin{equation}\label{eq:laplacebis1}
	\frac{d}{ds}\bigg( \mathbf n \cdot \pmb \sigma \cdot \mathbf n \bigg)_{\phi=0} = \gamma \frac{d^2\theta}{ds^2},
\end{equation} 
where we used that the atmospheric pressure outside the fluid is constant. It should be noted that this expression is exact, but we lack an explicit expression for the left hand side. The central idea behind the generalized lubrication equation is that for small ${\rm Ca}$ the interface curvature is small, so that the gradient of normal stress along the interface to leading order can be computed from the solution of a perfectly straight corner. Given that $\tau_{\phi \phi}=0$ at $\phi=0$, and using that $ds \simeq dr$, we can rewrite \eqref{eq:laplacebis1} as 
\begin{equation}\label{eq:laplacebis}
	\left. -\frac{\partial p}{\partial r } \right|_{\phi=0} = \gamma \frac{d^2\theta}{ds^2},
\end{equation} 
and use the pressure gradient derived in \eqref{eq:dpdr}. The left hand side involves the radial coordinate $r$, which for weak curvature $r \simeq h/\sin \theta$. With this, \eqref{eq:laplacebis} becomes
\begin{equation}
	\gamma \frac{d^2\theta}{ds^2} =
	\frac{2 \eta U}{h^2} \left(\frac{\sin^3{\theta}}{\theta-\cos{\theta}\sin{\theta}}\right) + \frac{2 \psi_1 U^2}{h^3} \left(\frac{\sin^4{\theta} (\sin{\theta} - \theta \cos{\theta})}{\left(\theta-\cos{\theta}\sin{\theta}\right)^2}\right).
\end{equation}
Combined with the geometric relation

\begin{equation}
	\frac{\partial h}{\partial s} = \sin \theta,
\end{equation}
this comprises a third order system for $\theta(s),h(s)$, defining the interface shape. 

In what follows we will be interested in the the dip-coating geometry, where a plate is withdrawn from or plunged into a bath with an angle $\theta_p$, as sketched in figure~\ref{fig:fig1}. This geometry requires the introduction of gravity as a body force, which can be easily incorporated into the description. Following the Newtonian case of~\citet{chan2013hydrodynamics}, the generalized lubrication equation takes the form
\begin{equation}\label{eq:GLwithoutslip}
	\gamma \frac{d^2\theta}{ds^2} = \frac{3 \eta U}{h^2} f(\theta) \left(1 + \frac{\psi_1}{\eta} \frac{U}{h} g(\theta) \right) + \rho g \sin{\left(\theta -\theta_p \right)},
\end{equation}
where for notational convenience we have collected the $\theta$-dependence associated to the corner flow in the two auxiliary functions $f(\theta)$ and $g(\theta)$ defined by

\begin{equation}
	f(\theta)= \frac{2}{3} \left(\frac{\sin^3{\theta}}{\theta-\cos{\theta}\sin{\theta}} \right) \quad \mathrm{and} \quad g(\theta) = \left( \frac{\sin^2{\theta}-\theta \cos{\theta}\sin{\theta}}{\theta - \cos{\theta}\sin{\theta}}\right).
\end{equation}
Setting $\psi_1=0$ in \eqref{eq:GLwithoutslip} we recover the generalized lubrication equation for a Newtonian fluid, with a no-slip boundary condition. The extra term involving $\psi_1$ represents the influence of the normal stress effect.  

\subsection{Lubrication limit and introducing slip} 
\label{sec:lubrication_limit_and_introducing_slip}

Before proceeding, we will analyse \eqref{eq:GLwithoutslip} in the limit $\theta \ll 1$. This serves two purposes. First, we will verify that the generalized lubrication theory reduces to the ``normal" lubrication theory, obtained via a long-wave expansion of the second-order fluid. Second, it enables to phenomenologically introduce a Navier slip condition; the corner solutions that form the basis of the generalized lubrication equation assumed a no-slip boundary condition at the solid, but this restriction needs to be lifted in order to regularise the moving contact line singularity.  

In the limit $\theta \ll 1$, we can make use of the limiting behaviour,
\begin{equation}
	s \simeq x, \quad \theta \simeq h', \quad f(\theta) \simeq 1 \quad \mathrm{and} \quad g(\theta) \simeq \frac{\theta}{2} \simeq \frac{h'}{2},
\end{equation}
so that \eqref{eq:GLwithoutslip} reduces to 
\begin{equation}\label{eq:smalltheta}
	\gamma h''' = \frac{3 \eta U}{h^2} + \frac{3}{2}\frac{\psi_1 U^2 h'}{h^3}  + \rho g \left( h' - \theta_p \right),
\end{equation}
where the plate angle has also been considered small, for $ \sin{\left(\theta -\theta_p \right)} \simeq h' -\theta_p$. This result is indeed perfectly consistent with the long-wave expansion obtained from the second-order fluid derived in \citet{datt2022thin}, which is given by
\begin{equation} \label{eqn:lubrication}
	\gamma h''' = \frac{3 \eta U}{h \left( h+ 3 \lambda_s \right)} + \frac{3}{2}\frac{\psi_1 U^2 h'}{\left( h+ 3 \lambda_s \right)^3}  + \rho g \left( h' - \theta_p \right),
\end{equation}
where $\lambda_s$ is the slip length. Indeed, setting $\lambda_s=0$, this result is identical to the small angle limit of the generalized lubrication equation without slip, as given in \eqref{eq:smalltheta}. To introduce an effective slip in the generalized \eqref{eq:GLwithoutslip}, we follow~\citet{chan2013hydrodynamics} and use the regularisation forms observed in the lubrication limit. This amounts to replacing $h^2$ by $h(h + 3\lambda_s)$ for the viscous term, and $h^3$ by $(h+3\lambda_s)^3$ for the viscoelastic term. For a detailed discussion on this approximation, we refer to~\citet{chan2020cox}, where the prefactor of the slip term has been computed for arbitrary contact angles in corner flows. It was shown by~\citet{chan2020cox} that the prefactor `3' for the slip term is strictly only valid in the limit of small angles. The more appropriate form is to replace $3\lambda_s$ by $k(\theta_e)\lambda_s$, where $k$ is a decreasing function of $\theta_e$. However, this function is not available in closed form, so that, for simplicity, we stick to the factor `3' throughout. This does not affect the general nature of the problem; for a given contact angle, one can correct for this effect by renormalizing the reported value of $\lambda_s/\ell_\gamma$ by a factor $3/k$.

\subsection{Dimensionless form (with slip) and boundary conditions} 
\label{sec:dimensionless_form}

We conclude this section by posing the equations in dimensionless form, including the phenomenological slip, and by discussing the boundary conditions that define the problem. Besides the microscopic slip length $\lambda_s$, the equations present two further length scales, 

\begin{equation}
	\ell_\gamma = \sqrt{\frac{\gamma}{\rho g}}, \quad \ell_{VE}= \frac{\psi_1 U}{\eta}.
\end{equation}
The balance of gravity and surface tension gives the capillary length $\ell_\gamma$, serving as the outer length scale of the problem that is orders of magnitude larger than $\lambda_s$ in practice. The balance of normal stress and viscosity gives rise to a viscoelastic length $\ell_{VE}$, which we vary from 0 (no elasticity) to values in the range $\lambda_s \ll \ell_{VE} \ll \ell_\gamma$.

In what follows, we choose to make the equations dimensionless using $\ell_\gamma$ as the relevant length scale. The generalized lubrication equation then becomes:
\begin{equation}
	\label{eqn:generalized_lubrication}
	\frac{\partial^2 \theta}{\partial \bar s^2} =  3 \mathrm{Ca} \frac{1}{\bar{h} \left( \bar{h} + 3 \bar \lambda_s \right)} f(\theta)  + 3 \mathrm{Ca} \bar{\ell}_{VE} \frac{1}{\left( \bar{h} + 3 \bar \lambda_s \right)^3} f(\theta) g(\theta)  + \sin{\left(\theta -\theta_p \right)}, 
\end{equation} 
where symbols with an overbar are lengths scaled by $\ell_\gamma$. In this equation we further defined the capillary number as ${\rm Ca}=\eta U/\gamma$, which is positive (negative) for receding (advancing) contact line motion. This equation needs to be complemented by the geometric relation
\begin{equation}
	\label{eqn:dh_ds}
	\frac{d\bar h}{d\bar s} = \sin \theta.
\end{equation}
The boundary conditions for this third order system are 

\begin{equation} \label{eqn:BCs}
	\bar h(\bar s=0)=0, \quad \theta(\bar s=0)=\theta_e, \quad \theta(\bar s\to \infty) = \theta_p,
\end{equation}
which impose the partial wetting contact line condition at $\bar s=0$ and the connection to the bath as $\bar s\to \infty$. 

We numerically solve the problem by using the continuation code AUTO-07P~\citep{doedel2007auto}. For Newtonian fluids, the slip length is nanometric, while the capillary length is of the order of a millimeter. The slip length in viscoelastic fluid is less known and can largely exceed the nanometer range~\citep{guyard2021near}. To have a good scale separation between slip and capillary length, we choose $\bar \lambda_s=10^{-4}$ that is left constant in the rest of the paper. We point out that numerically it has been a challenge to use dimensionless slip length below $\bar \lambda_s=10^{-4}$ in the presence of strong viscoelastic effects. The viscoelastic numerical solver always breaks beyond a certain absolute value of the capillary number, where no solutions are found. To validate our numerical results, we also used a different method to solve \eqref{eqn:generalized_lubrication}, namely the shooting method. Both solvers give identical results, including near the point where the codes breakdown. The $\rm Ca$ value at which solutions of \eqref{eqn:generalized_lubrication} can no longer be integrated numerically is non-trivial and depends both on $\bar{\ell}_{VE}$ and $\bar{\lambda}_s$. This numerical issue does not seem to be related to any physical processes as the numerical interface profiles are very smooth near the breaking point. Further investigations would be necessary to understand this break down. 

The problem defined above thus contains 5 parameters, namely

\begin{equation}
	{\rm Ca} = \frac{\eta U}{\gamma}, \quad \bar{\ell}_{VE}=\frac{\ell_{VE}}{\ell_\gamma}, \quad \bar \lambda_s = \frac{\lambda_s}{\ell_\gamma}, \quad \theta_e, \quad \theta_p.
\end{equation}
In comparison to the Newtonian case, there is an additional viscoelastic parameter that can be identified as a Weissenberg number
\begin{equation}
	\mathrm{Wi}=  \bar{\ell}_{VE} =  \frac{\ell_{VE}}{\ell_\gamma}= \frac{\psi_1 U}{\eta \ell_\gamma}.
\end{equation}
This number compares the intrinsic viscoelastic time $\psi_1/\eta$ to the shear rate $U/\ell_\gamma$ based on the capillary length. Choosing $\rm Wi$ as a viscoelastic parameter does not facilitate direct comparisons with experimental results, as the contact line speed $U$ (which is a control parameter) appears in both $\rm Ca$ and $\rm Wi$. The lubrication analysis presented in \citet{kansal2024viscoelastic}, revealed the importance of another dimensionless parameter, namely  

\begin{equation}
	N_0=  \frac{\psi_1 \gamma \theta_e^4}{\eta^2 \lambda_s}= \frac{\mathrm{Wi}}{\mathrm{Ca}} \frac{\theta_e^4}{\bar \lambda_s}. 
\end{equation}
Conveniently, this dimensionless number scales the normal stress coefficient $\psi_1$ with only material parameters, and hence it can be seen as a material property. Most of the results presented below will therefore be presented using $N_0$ as the relevant viscoelastic parameter, such that ${\rm Ca}$ can be interpreted as the dimensionless control parameter for the plate speed at fixed liquid properties. 

\section{Dipcoating results}
\label{sec:application_to_dip_coating}

\subsection{Phenomenology} 
\label{sec:phenomenology}

\begin{figure}
	\centerline{\includegraphics[width=\linewidth]{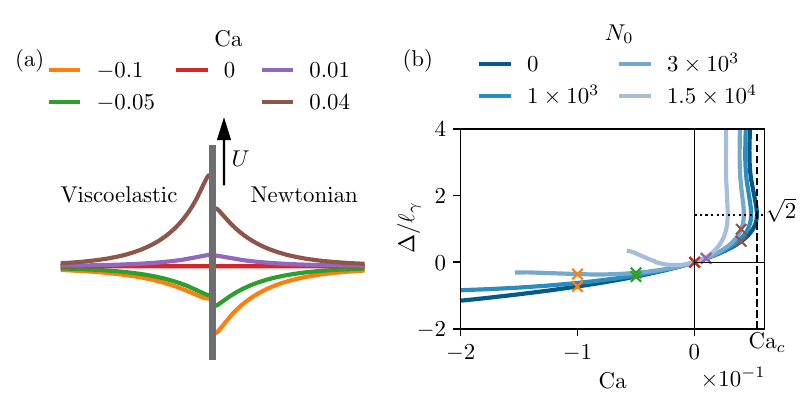}}
	\caption{Phenomenology of viscoelastic dip-coating obtained by numerically integrating equation~\eqref{eqn:generalized_lubrication}, illustrated for $\theta_p =90^\circ, \theta_e=90^\circ$, and $\lambda_s/\ell_\gamma=10^{-4}$. (a) Typical interface profiles for advancing motion (${\rm Ca}<0$, green and orange curves) and receding motion (${\rm Ca}>0$, brown and purple curves). The right side displays the Newtonian solutions for $N_0=0$, while the viscoelastic solutions (left) are plotted for $N_0=1 \times 10^3$. (b) Vertical position of the contact line $\bar \Delta = \Delta/\ell_\gamma$ versus capillary number $\mathrm{Ca}$, for different values of the viscoelastic material parameter $N_0$. The crosses correspond to the interface profiles of figure (a) using the same color code. The vertical dashed line represents the critical capillary number for $N_0=0$, beyond which no solution exists as a Landau-Levich film is entrained. The solutions in the upper branch beyond $\Delta/\ell_\gamma = \sqrt{2}$ will not be discussed in this work. In the viscoelastic case, the numerical integration of \eqref{eqn:generalized_lubrication} cannot be achieved for Ca below a certain value, as discussed in section~\ref{sec:advancing_contact_lines}. }
	\label{fig:fig_overall_vertical_plate}
\end{figure}

\begin{figure}
	\centerline{\includegraphics[width=\linewidth]{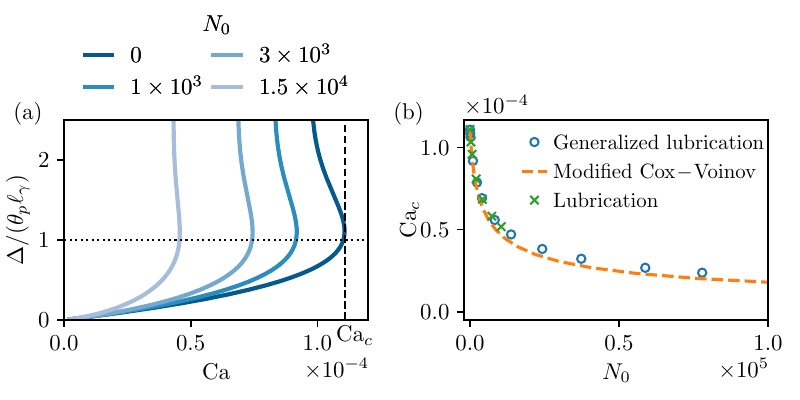}}
	\caption{Receding contact lines, for small angles $\theta_p =10^\circ, \theta_e=10^\circ$, and $\lambda_s/\ell_\gamma=10^{-4}$. (a) Contact line position  $\bar \Delta/\theta_p=\Delta/\left( \theta_p \ell_\gamma \right)$ versus capillary number $\mathrm{Ca}$. (b) Critical capillary number $\mathrm{Ca}_c$ as a function of the viscoelastic material parameter $N_0$. The blue circles are obtained by numerically integrating generalized lubrication equation~\eqref{eqn:generalized_lubrication}. The green crosses are the numerical solutions of the lubrication equation~\eqref{eqn:lubrication}. The modified Cox-Voinov represents the asymptotic expansion~\eqref{eq:viscoelastic_wetting_transition} using~\eqref{eqn:c3_small_angles}. }
\label{fig:fig_receding_small_plate_angle}
\end{figure}

We start by showing typical results obtained from the numerical integration of the generalized lubrication equation~\eqref{eqn:generalized_lubrication}. Throughout this paper the slip length is kept fixed at $\bar \lambda_s=10^{-4}$, offering a good compromise between accurate numerical resolution and sufficient separation of the microscopic and macroscopic scales of the problem (also discussed in section~\ref{sec:dimensionless_form}). We illustrate the phenomenology for the case where the plate is vertical, $\theta_p=90^\circ$, and for simplicity we consider $\theta_e=90^\circ$. This choice of $\theta_e=90^\circ$ is well-suited for the illustration as it makes the equilibrium configuration symmetric in terms of wetting, and as a consequence the equilibrium state consists of a horizontal meniscus in contact with a vertical plate (cf. figure~\ref{fig:fig_overall_vertical_plate}(a), at ${\rm Ca=0}$). As the plate speed is increased, the meniscus is drawn up (pulling plate) or pushed down (plunging plate), leading to a rise or depression of the contact line position that is quantified by $\bar \Delta$ ($=\Delta/\ell_\gamma$, cf. figure~\ref{fig:fig1}). We remark that we also consider the case of tilted plate in further sections, with the plate angle $\theta_p=10^\circ$ both for advancing and receding contact lines (see figures~\ref{fig:fig_receding_small_plate_angle}, \ref{fig:fig_adv_small_angles} and \ref{fig:fig_adv_small_angles_with_Boudaoud}).

Figure~\ref{fig:fig_overall_vertical_plate}(b) reports the meniscus height $\bar \Delta$ as a function of dimensionless plate speed ${\rm Ca}$, for different strengths of viscoelasticity characterised by $N_0$. Let us first focus on the Newtonian case $N_0=0$, for which no viscoelasticity is present. As can be seen, despite the symmetric wetting conditions, the dip-coating is very asymmetric depending on whether the contact line is advancing or receding~\citep{chan2013hydrodynamics}. In the advancing case (${\rm Ca}<0$), increasing the plate speed leads to a deeper meniscus -- a process that can be continued to arbitrary ${\rm Ca}$ as long as no outer fluid is taken into account in the description~\citep{lorenceau2004air,marchand2012air}. For receding contact lines (${\rm Ca}>0$), however, steady solutions only exist up to a critical speed ${\rm Ca}_c$ beyond which a liquid film is entrained~\citep{blake1979maximum,sedev1991critical,eggers2005existence,snoeijer2007relaxation,chan2012theory,galvagno2014continuous}. As can be seen, the critical point is close to $\bar \Delta = \sqrt{2}$. This result has a natural interpretation in terms of a vanishing apparent 
macroscopic contact angle. Namely, the height of the static outer solution is given by~\citep{gennes2004capillarity}

\begin{equation}\label{eq:rise_eq}
	\bar \Delta 
	= \pm \sqrt{ 2 \left( 1 - \cos(\theta_p - \theta_{{\rm app},o} )\right)},
\end{equation}
where $\theta_{{\rm app},o}$ is the apparent ``outer" angle of the meniscus measured with respect to the plate, which reduces to the equilibrium contact angle $\theta_{e}$ for the static case. The sign in this expression is positive (negative) when $\theta_p > \theta_{{\rm app},o}$ ($\theta_p < \theta_{{\rm app},o}$). The maximum rise of the meniscus is achieved when $\theta_{{\rm app},o}=0$, which indeed explains the observed height at the critical point in figure~\ref{fig:fig_overall_vertical_plate}(b): for a vertical plate $\theta_p=90^\circ$ and a vanishing apparent angle implies $\bar \Delta=\sqrt{2}$. Note that dynamical solutions exist above $\bar \Delta=\sqrt{2}$; clearly, such solutions do not admit an interpretation based on an apparent macroscopic angle as the maximum $\bar \Delta$ from equation~\eqref{eq:rise_eq} is $\sqrt{2}$~\citep{chan2012theory}. 

Our main interest, however, is to explore the effect of viscoelasticity on these results, which is achieved via the variation of $N_0$. Focusing on the receding side of figure~\ref{fig:fig_overall_vertical_plate}(b), one observes that ${\rm Ca}_c$ becomes smaller upon increasing $N_0$. This implies that viscoelasticity facilitates the onset of liquid entrainment, as entrainment appears at smaller plate velocities. The height of the meniscus at the critical point, however, remains close to $\bar \Delta=\sqrt{2}$, such that $\theta_{{\rm app},o}$ is still equal to $0$ as in Newtonian case. Focussing next on the advancing side of figure~\ref{fig:fig_overall_vertical_plate}(b), we observe that viscoelasticity leads to an increase of $\bar \Delta$ with respect to the Newtonian case ($N_0=0$). For sufficiently large $N_0$, this can even lead to the emergence of a minimum in $\bar \Delta$, which is a remarkable result: in this regime, pushing the plate at higher speed into the bath leads to a rise of the meniscus. 

We thus conclude that for both advancing and receding contact lines, the effect of viscoelasticity is to give a higher $\bar \Delta$ as compared to the Newtonian case. Using \eqref{eq:rise_eq}, this points to a lowering of the apparent macroscopic angle induced by viscoelasticity. In the remainder we wish to explore these findings in more detail, and provide quantitative explanations for these observations. 

\subsection{Receding contact lines} 
\label{sec:receding}

\subsubsection{Small angles}

The first question we wish to address systematically is how ${\rm Ca}_c$, the critical receding speed for liquid entrainment, is affected by viscoelasticity. To establish connection with the results previously derived in the context of lubrication theory \citep{kansal2024viscoelastic}, we consider a case for which $\theta_p=\theta_e = 10^\circ$, ensuring that interface slopes $\theta$ remain small on the entire domain, as $\theta \to \theta_p$ while approaching the liquid bath~\eqref{eqn:BCs}. The small plate angle also ensures $\left( \theta -\theta_p \right)$ to be small, as it was assumed for simplification of the gravity term in~\eqref{eq:smalltheta} and subsequently in the viscoelastic lubrication equation~\eqref{eqn:lubrication}. The result is shown in figure~\ref{fig:fig_receding_small_plate_angle}. Panel (a) shows the rise of the meniscus as a function of ${\rm Ca}$, for different values of $N_0$. The phenomenology is the same as in figure~\ref{fig:fig_overall_vertical_plate}, but now the critical point arises close to $\bar \Delta/\theta_p =1$. Using the small-angle approximation of \eqref{eq:rise_eq}, this once again coincides with a vanishing $\theta_{{\rm app},o}$. The dependence of ${\rm Ca}_c$ on $N_0$ is shown in figure~\ref{fig:fig_receding_small_plate_angle}(b). The circles show the results from the numerical solutions of the generalized lubrication equation \eqref{eqn:generalized_lubrication}. As expected, these perfectly match the crosses obtained from numerical integration of the viscoelastic lubrication equation~\eqref{eqn:lubrication}. 

\citet{kansal2024viscoelastic} derived an analytical prediction for the critical speed in the lubrication limit with viscoelastic effects by using the asymptotic matching method. The normal stress term in \eqref{eqn:generalized_lubrication} scales as $h^{-3}$ and is more singular than the viscous term $\propto h^{-2}$ towards the contact line ($h\to 0$). Hence, the action of the normal stress effects is localised near the contact line, at length scales $\sim \lambda_s$. However, for length scales larger than the slip length, the viscoelasticity does not affect the global balance, and the contact line motion can be understood in the framework of a modified Cox-Voinov theory. More precisely, the inner asymptotic of the slope angle follows  
\begin{equation}
\label{eq:modified-cox-voinov}
\theta^3(x) = \theta_{\mathrm{app},i}^3 - 9 \,\mathrm{Ca} \ln\left( \frac{e x \theta_e}{3\lambda_s}\right), \quad \quad \theta_{\mathrm{app},i}^3 = \theta_e^3 - \frac{3}{4}\frac{\psi_1 U^2 \theta_e}{\gamma \lambda_s},
\end{equation}
for $\lambda_s/\theta_e \ll x \ll \ell_\gamma$, where $e = \exp(1)$ and $x$ the distance from the contact line (see Fig.~\ref{fig:fig1}). We introduce $\theta_{\mathrm{app},i}$ in \eqref{eq:modified-cox-voinov} which is the so-called apparent inner angle. The effects of viscoelasticity on the contact line motion is to modify the microscopic boundary condition of the Cox-Voinov relation and to decrease the inner angle, ultimately leading to a decrease of the macroscopic outer angle $\theta_{\mathrm{app},o}$. We anticipate that the same qualitative picture also holds for advancing contact line and in the weak viscoelastic limit as discussed below. Following the procedure in \citet{eggers2005existence}, \citet{kansal2024viscoelastic} matched the modified Cox-Voinov expression~\eqref{eq:modified-cox-voinov} to the static bath solution at vanishing apparent angle in order to find the critical speed. Using dimensionless variables, the critical receding speed for liquid entrainment follows

\begin{equation}
\label{eq:viscoelastic_wetting_transition}
\frac{\mathrm{Ca_c}}{\theta_e^3} = \frac{1-\frac{3}{4} N_0 \left(\frac{\mathrm{Ca_c}}{\theta_e^3}\right)^2}{9 \ln \left(c_3 \left(\frac{\mathrm{Ca_c}}{\theta_e^3}\right)^{1/3} \right)},
\end{equation}
where 
\begin{equation} \label{eqn:c3_small_angles}
c_3 = \frac{1}{18^{1/3} \pi [\mathrm{Ai}(s_\mathrm{max})]^2} \frac{\ell_\gamma \theta_e^2}{\lambda_s \theta_p} \approx 0.423 \frac{\theta_e^2}{\bar \lambda_s \theta_p}.
\end{equation}
This analytical prediction is superimposed as the dashed line in figure~\ref{fig:fig_receding_small_plate_angle}(b), in excellent agreement with the numerical model.

\subsubsection{Large angles}

\begin{figure}
\centerline{\includegraphics[width=\linewidth]{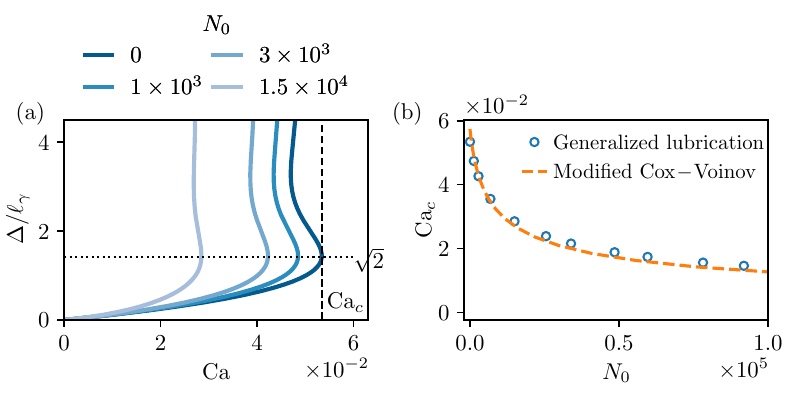}}
\caption{Receding contact lines, for $\theta_p =90^\circ, \theta_e=90^\circ$, and $\lambda_s/\ell_\gamma=10^{-4}$. (a) Contact line position  $\bar \Delta = \Delta/ \ell_\gamma$ versus capillary number $\mathrm{Ca}$. (b) Critical capillary number $\mathrm{Ca}_c$ as a function of the viscoelastic material parameter $N_0$. Similarly to figure~\ref{fig:fig_receding_small_plate_angle}, blue circles are numerical solutions of~\eqref{eqn:generalized_lubrication}, while the orange dashed line represents the modified Cox-Voinov prediction corresponding to~\eqref{eq:viscoelastic_wetting_transition} using~\eqref{eqn:c3_large_angles}.}
\label{fig:fig_rec_vertical_plate_fixed_contact_angle_90}
\end{figure}	

We now extend the quantitative analysis of ${\rm Ca}_c$ to the case of large angles. Figure~\ref{fig:fig_rec_vertical_plate_fixed_contact_angle_90}(a) reports a detailed view of the case where $\theta_p=\theta_e=90^\circ$ (same data as figure~\ref{fig:fig_overall_vertical_plate}). The relation between ${\rm Ca}_c$ and the viscoelastic parameter $N_0$ is reported in figure~\ref{fig:fig_rec_vertical_plate_fixed_contact_angle_90}(b), and exhibits a dependency that closely resembles that observed in the lubrication limit. We therefore take a bold step and consider the analytical result obtained in the lubrication approximation, and superimpose it onto the numerical data for large angles as the dashed line. Specifically, we use the exact same form \eqref{eq:viscoelastic_wetting_transition}, but with a modified parameter $c_3$. Namely, the value of $c_3$ given in \eqref{eqn:c3_small_angles} emerges from matching the inner asymptotic expression to the curvature of the outer solution, which consists of a static bath with vanishing apparent angle: for small plate inclinations this curvature is equal to $\theta_p$, while it is $\sqrt{2}$ for a vertical plate (expressed in units of capillary length) \citep{bonn2009wetting}. By consequence, we use 
\begin{equation} \label{eqn:c3_large_angles}
c_3 = \frac{1}{18^{1/3} \pi [\mathrm{Ai}(s_\mathrm{max})]^2} \frac{\ell_\gamma \theta_e^2}{\lambda_s \sqrt{2}} \approx 0.423 \frac{ \theta_e^2}{\bar \lambda_s \sqrt{2}},
\end{equation}
for a vertical plate. As is clear from figure~\ref{fig:fig_rec_vertical_plate_fixed_contact_angle_90}(b), this expansion provides an excellent description of the critical speed, despite the fact that we are encountering interface slopes as large as $\theta_e = 90^\circ$. 

A few remarks are in order here. The fact that lubrication results carry over to large angles is well-known for Newtonian wetting. The large-angle expansion by \citet{cox_1986} involves an integral expression that is almost equal to $\theta^3$, with differences of a few percent for angles as large as $150^\circ$ \citep{snoeijer2013moving}. This property offers a rationalisation for the fact that the lubrication result is accurate for the Newtonian limit $N_0=0$. However, the remarkable feature evidenced by figure~\ref{fig:fig_rec_vertical_plate_fixed_contact_angle_90}(b) (where $\theta_e = 90^\circ$) is that the dependence on $N_0>0$ is also accurately captured by lubrication result \eqref{eq:viscoelastic_wetting_transition}. This result deserves further analytical motivation in future analysis. As discussed at the end of section \ref{sec:lubrication_limit_and_introducing_slip}, we recall that for large contact angles the parameter $\lambda_s$ is an effective slip length that is not exactly equal to (yet proportional to) the Navier-slip length -- this proportionality factor needs to be taken into account in \eqref{eqn:c3_large_angles} when attempting a fully quantitative description. 

\begin{figure}
\centerline{\includegraphics[width=\linewidth]{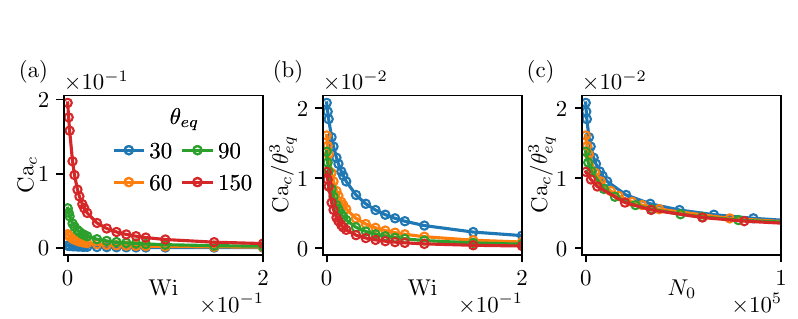}}
\caption{Receding contact lines with varying wettability of the plate, $\theta_e=30^\circ, 60^\circ, 90^\circ \& \, 150^\circ$, for fixed values $\lambda_s/\ell_\gamma=10^{-4}, \theta_p =90^\circ$. (a) Critical capillary number $\mathrm{Ca}_c$ versus Weissenberg number $\mathrm{Wi}$. (b) Rescaled critical capillary number $\mathrm{Ca}_c/\theta_e^3$ versus Weissenberg number $\mathrm{Wi}$. (c) Rescaled critical capillary number $\mathrm{Ca}_c/\theta_e^3$ versus viscoelastic material parameter $N_0$.}
\label{fig:fig_rec_vertical_plate_varying_contact_angle}
\end{figure}	

We conclude the discussion of receding contact lines by systematically varying $\theta_e$ for a vertical plate, and try to further confirm the scaling laws for the critical speed. Figure~\ref{fig:fig_rec_vertical_plate_varying_contact_angle}(a) shows the corresponding ${\rm Ca}_c$, now plotted as a function of the Weissenberg number ${\rm Wi}$. We attempt to collapse the data, first by the rescaling ${\rm Ca}_c/\theta_e^3$. Figure~\ref{fig:fig_rec_vertical_plate_varying_contact_angle}(b) shows that this rescaling brings the data closer in the vertical direction, but the dependence on ${\rm Wi}$ is clearly not universal. A much better collapse is indeed achieved in figure~\ref{fig:fig_rec_vertical_plate_varying_contact_angle}(c), when using the viscoelastic parameter $N_0$ rather than ${\rm Wi}$. This confirms that $N_0$ offers the correct scaling of the normal stress effect on receding contact line motion, as expressed by \eqref{eq:viscoelastic_wetting_transition}. Note that upon close inspection of figure~\ref{fig:fig_rec_vertical_plate_varying_contact_angle}(c), one observes that the numerical data do not perfectly collapse. Indeed, as can be seen in equations~\eqref{eq:viscoelastic_wetting_transition} and~\eqref{eqn:c3_small_angles}, the coefficient $c_3$ exhibits a dependence $\theta_e^2$, leading to non-universal logarithmic corrections that prevent a perfect collapse of the data.

\subsection{Advancing contact lines}
\label{sec:advancing_contact_lines}

\begin{figure}
\centerline{\includegraphics[width=\linewidth]{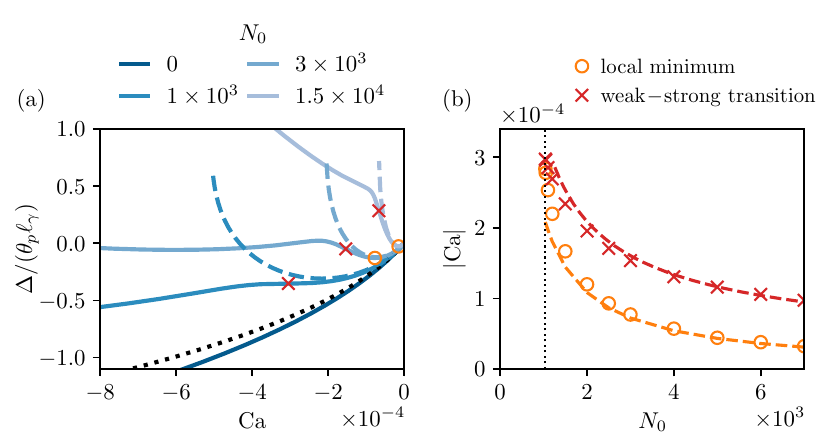}}
\caption{Advancing contact lines, for small angles $\theta_p =10^\circ, \theta_e=10^\circ$, and $\lambda_s/\ell_\gamma=10^{-4}$. (a)  Contact line position $\bar \Delta/\theta_p=\Delta/\left( \theta_p \ell_\gamma \right)$ versus capillary number $\mathrm{Ca}$. The local minima and the inflection points are represented by orange circles and red crosses respectively. The dashed lines represent the predictions for the weakly viscoelastic regime~\eqref{eq:CVplung}, using the relation between $\Delta/(\theta_p \ell_\gamma)$ and $\theta_{\mathrm{app},o}$ from~\eqref{eq:Delta_thetap_minus_thetaapp}. (b) Absolute values of the capillary number at the local minimum of $\bar \Delta/\theta_p$ and at the inflection points, plotted as a function of the viscoelastic material parameter $N_0$. The dashed lines are the asymptotic predictions, given by~\eqref{eq:local_minimum_advancing_Delta} and~\eqref{eq:advancing_transition}.}
\label{fig:fig_adv_small_angles}
\end{figure}	

We start the discussion of advancing contact lines with a case where $\theta_p=\theta_e \ll 1$, to enable a comparison to lubrication theory. Figure~\ref{fig:fig_adv_small_angles}(a) shows the meniscus indentation ($\bar \Delta <0$) as a function of plunging speed (${\rm Ca}<0$). While for $N_0=0$, the decrease of $\bar \Delta$ is monotonic with plate speed, this is no longer true for sufficiently large normal stress. Instead, a local minimum appears and the curves also exhibit a kink suggesting the transition toward a new regime. At large $N_0$ (typically $3 \times 10^3$ in figure~\ref{fig:fig_adv_small_angles}(a)), the local minimum also becomes the global minimum of the meniscus indentation which increases at large $|{\rm Ca}|$. These features are further reported in figure~\ref{fig:fig_adv_small_angles}(b), showing the value of the capillary number at the local minimum (orange circles), which appears above some critical value of $N_0$, around $10^3$. We also report the capillary number at the inflection point of the curve (red crosses), to give an idea of the location of the kink. While the values of ${\rm Ca}$ of the minimum and the inflection point are initially close, they exhibit a different dependence on $N_0$.

Once again, these features can be understood from a theoretical analysis of the lubrication equation, as developed in \citet{kansal2024viscoelastic}. Specifically, for advancing contact lines \citet{kansal2024viscoelastic} derived two modified forms of the Cox-Voinov law for the apparent macroscopic angle. In the limit of weak viscoelasticity, the modified Cox-Voinov expression~\eqref{eq:modified-cox-voinov} provides a very good description of the interface slope inner solution. We follow~\citet{eggers2005existence} and match the interface slope to a static bath with an apparent angle $\theta_{\mathrm{app},o}$, which leads to
\begin{equation}\label{eq:CVplung}
\theta_{\mathrm{app},o}^3 = \theta_{\mathrm{app},i}^3 + 9 |\mathrm{Ca}| \ln{\left( \frac{e \ell_\gamma \theta_e b}{3 \lambda_s}\right)}, \quad \quad b= e^{-\Gamma}, 
\end{equation}
where $\Gamma$ is the Euler-Mascheroni constant. The numerical constant $b$ is found from the matching procedure and depends in a non trivial way in the ratio between plate angle and equilibrium angle \citep{eggers2005existence}, which in \eqref{eq:CVplung} is given for the specific value of $\theta_p = \theta_e$. The apparent macroscopic angle $\theta_{{\rm app},o}$ exhibits a non-monotonic dependence that can be understood from a competition of the viscous increase ($\sim U$) and the decrease of the inner angle due to normal stresses $(\sim U^2)$. In connection with the small-plate angle expansion of \eqref{eq:rise_eq}, 
\begin{equation} \label{eq:Delta_thetap_minus_thetaapp}
\bar \Delta = \theta_p - \theta_{\mathrm{app},o},
\end{equation}
which results into the dashed lines superimposed in figure~\ref{fig:fig_adv_small_angles}(a). As can be seen, this modified form of the Cox-Voinov theory indeed closely follows the data at small ${\rm Ca}$ and captures the emergence of a local minimum. The capillary number at the minimum can be evaluated from \eqref{eq:CVplung}, which in the present notation gives 
\begin{equation} \label{eq:local_minimum_advancing_Delta}
|{\rm Ca}| = 6 \ln{\left( \frac{e \theta_e b}{3 \bar{\lambda}_s}\right)}/N_0,
\end{equation}	
This prediction is superimposed in figure~\ref{fig:fig_adv_small_angles}(b) (orange dashed line) and provides a very good description of the local minimum of the meniscus indentation.

\begin{figure}
\centerline{\includegraphics[width=0.5\linewidth]{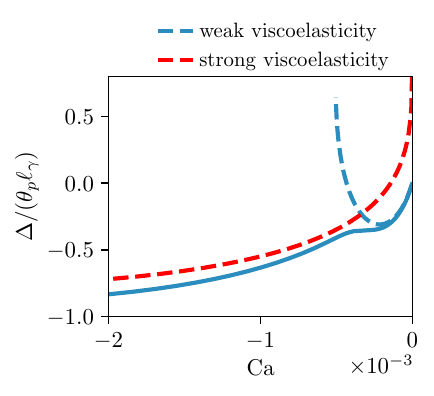}}
\caption{Advancing contact lines, for small angles $\theta_p =10^\circ, \theta_e=10^\circ$, and $\lambda_s/\ell_\gamma=10^{-4}$, for a selected value of $N_0=1 \times 10^3$. The two dashed lines represent the predictions for the weakly viscoelastic regime~\eqref{eq:CVplung} and the strongly viscoelastic regime~\eqref{eq:CVplungbis}, using the relation between $\Delta/(\theta_p \ell_\gamma)$ and $\theta_{\mathrm{app},o}$ from~\eqref{eq:Delta_thetap_minus_thetaapp}.}
\label{fig:fig_adv_small_angles_with_Boudaoud}
\end{figure}	

However, this analysis does not predict the onset of the kink. To describe the kink, we invoke the second viscoelastic advancing contact line law derived in \citet{kansal2024viscoelastic}. Indeed, for advancing contact lines in the strong viscoelasticity limit, the modified Cox-Voinov law of \eqref{eq:modified-cox-voinov} no longer holds. When normal stress effects dominate capillary forces, a region of ``apparent complete wetting'' (flat film of typical lateral extent $\sim \ell_{VE}$) emerges near the contact line for $\lambda_s/\theta_e \ll x \ll \ell_{VE}$. The interface slope inner asymptotic still follows a modified Cox-Voinov law, but is now independent of the equilibrium angle and is entirely governed by the viscoelastic length scale as $\theta^3(x) = 9 |\mathrm{Ca}| \ln{\left( 2 x/(\xi_0 \ell_{VE}) \right)}$, where $\xi_0 \approx 1.16$ is a universal numerical constant. The matching procedure to the static bath at a apparent outer angle $\theta_{\mathrm{app},o}$ leads to 
\begin{equation}\label{eq:CVplungbis}
\theta_{\mathrm{app},o}^3 = 9 |\mathrm{Ca}| \ln{\left( \frac{2 \ell_\gamma b}{\xi_0 \ell_{VE}}\right)}, \quad \quad \xi_0 \approx 1.16.
\end{equation}
To further illustrate how \eqref{eq:CVplung} and \eqref{eq:CVplungbis} relate to the numerical results, we replot some of the data and the predictions in figure~\ref{fig:fig_adv_small_angles_with_Boudaoud}. The former is indicated as the blue dashed line, and closely follows the numerical data at small ${|\rm Ca|}$, where viscoelastic effects are weak. The latter is indicated as the red dashed line, and closely follows the numerical data at large ${\rm Ca}$, where viscoelastic effects are strong. Clearly, the appearance of a kink can be seen as the crossover from weak-to-strong viscoelasticity. For this crossover, we can simply equate the right hand sides of the equations~\eqref{eq:CVplung} and~\eqref{eq:CVplungbis}, which gives
\begin{equation}
\theta_e^3 - \frac{3}{4} \frac{\psi U^2\theta_e}{\gamma \lambda_s} + 9 |\mathrm{Ca}| \ln{\left( \frac{e \ell_\gamma \theta_e b}{3 \lambda_s}\right)} = 9 |\mathrm{Ca}| \ln{\left( \frac{2 \ell_\gamma b}{\xi_0 \ell_{VE}}\right)}.
\end{equation}
Simplifying this expression by introducing $N_0$, the capillary number at the weak-strong viscoelastic transition is the solution of the equation 
\begin{equation}
\label{eq:advancing_transition}
\frac{|\mathrm{Ca}|}{\theta_e^3} = - \frac{1-\frac{3}{4} N_0 \left(\frac{|\mathrm{Ca}|}{\theta_e^3}\right)^2}{9 \ln \left(c_4 N_0 \left( \frac{|\mathrm{Ca}|}{\theta_e^3}\right) \right)},
\end{equation}
where $c_4 = e \xi_0/6$. This prediction is superimposed in figure~\ref{fig:fig_adv_small_angles}(b) (red dashed line). We point out that the latter expression resembles the critical entrainment speed~\eqref{eq:viscoelastic_wetting_transition} discussed above in the receding case, up to a modification of the argument in the logarithmic term. In the limit of large $N_0$, the latter expression reduces to $|{\rm Ca}| \sim 2\theta_e^3/\sqrt{3N_0}$, which can be interpreted as the $|\rm Ca|$ at which the apparent inner angle $\theta_{{\rm app},i}\to 0$, which was argued to be the criterion governing the transition from weak-to-strong elasticity in \citet{kansal2024viscoelastic}.

\begin{figure}
\centerline{\includegraphics[width=\linewidth]{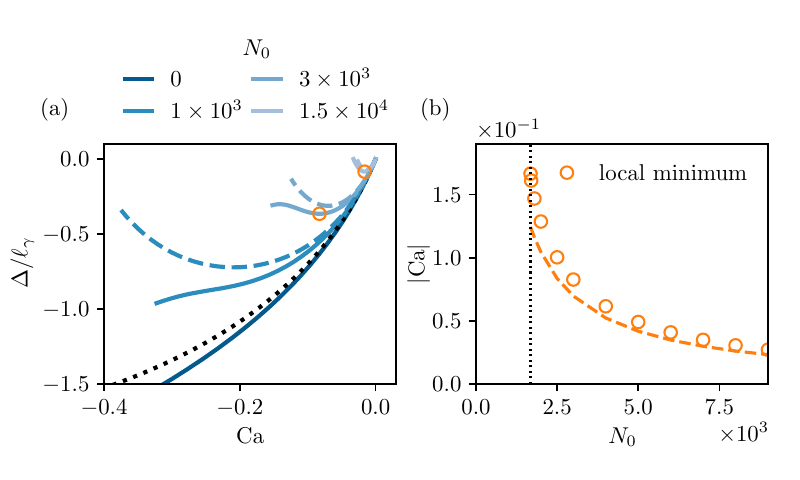}}
\caption{Advancing contact lines, for $\theta_p =90^\circ, \theta_e=90^\circ$, and $\lambda_s/\ell_\gamma=10^{-4}$. (a) Contact line position depth $\bar \Delta = \Delta/\ell_\gamma$ versus capillary number $\mathrm{Ca}$. (b) Capillary number at the location of local minimum of $\bar \Delta$ as a function of the viscoelastic material parameter $N_0$. The dashed lines in (a) and (b) represent the predictions for the weakly viscoelastic regime in~\eqref{eq:CVplung} and~\eqref{eq:local_minimum_advancing_Delta} respectively.}
\label{fig:fig_adv_vertical_plate}
\end{figure}
We conclude by figure~\ref{fig:fig_adv_vertical_plate}, where we discuss the same features for a case of large angles,  $\theta_e = \theta_p =90^\circ$. We also superimpose the meniscus position obtained using the modified Cox-Voinov law in the weak viscoelastic limit~\eqref{eq:CVplung}. Once again, we observe the appearance of a local minimum that is well-captured by the Cox-Voinov model~\eqref{eq:local_minimum_advancing_Delta}. The numerical data also show hint of a second regime, including the kink. Unfortunately, however, the numerical scheme was unable to produce quantitative data in this regime, which at present inhibits further quantitative analysis.

\section{Conclusion}
\label{sec:conclusion}
In this paper, we derived the generalized lubrication equation for the second-order fluid, which offers a model to study contact line motion with normal stress at arbitrary contact angles. As was the case for the generalized lubrication theory for Newtonian fluids, this approach is based on a perturbation expansion around exact corner flow solutions. Owing to Tanner's theorem \citep{tanner1966plane}, this involves the same corner flow profiles as for the viscous case \citep{huh1971hydrodynamic}. However, the momentum balance involves extra terms that are not present for Newtonian flows, and which represent the normal stresses that arise in viscoelastic fluids.

The generalized lubrication theory was used to study contact line motion for the dip-coating geometry at arbitrary plate angles and contact angles. Both receding and advancing contact lines were studied by pulling and pushing the plate from/into the liquid bath respectively. For the receding case, the lubrication result of \citet{kansal2024viscoelastic} for the critical capillary number for film entrainment as a function of material viscoelastic parameter~\eqref{eq:viscoelastic_wetting_transition} were validated for small angles. Interestingly, the analytical lubrication predictions are remarkably accurate even for large angles, as was oberved for a vertical plate and contact angles as high as $150^\circ$. In figure~\ref{fig:fig_rec_vertical_plate_varying_contact_angle}(c), we observe a nearly perfect collapse of the data using the same viscoelastic dimensionless parameter as for the viscoelastic lubrication theory. Predictions from the Newtonian lubrication theory are known to be surprisingly accurate for contact line motion at large angles, which can be understood from the analysis by \citet{cox_1986}. However, why the same holds true for the second-order fluid at large angles is not obvious a priori, and is a point that deserves further analysis.  

We then extended the analysis for advancing contact lines using the plunging dip-coating geometry. The advancing viscoelastic contact line motion follows a crossover from a weak to strong viscoelastic regimes, as predicted asymptotically in the small slope limit with the lubrication theory~\citep{kansal2024viscoelastic}. This crossover is intricate as it can involve a local minimum of the meniscus position. The plate speed at which this minimum is reached can be predicted analytically from lubrication theory, and once again the result appears to be quantitatively accurate also for large contact angles. However, the numerical findings could not resolve the strongly viscoelastic regime for large angles. Further analysis is required to see if the corresponding form of the Cox-Voinov law, as given by~\eqref{eq:CVplungbis}, still remains valid for large contact angles. 

It is of interest to compare our theory to that of \citet{han2014theoretical}, which is also based on corner solution for the second-order fluid. That work directly used the pressure (rather than the pressure gradient) of the corner solutions to evaluate the local curvature. As a consequence, \citet{han2014theoretical} predict that the curvature goes to zero at large distance from the contact line -- this is incorrect for receding contact lines, which requires the interface to match the bath with a finite curvature \citep{eggers2005existence}. Hence, the description of receding contact lines calls for the approach developed in the present paper, based on gradients of pressure. Having said that, for advancing contact lines the correct solution does exhibit a vanishing curvature at large distance. Indeed, for advancing contact lines in the weakly viscoelastic regime, our equations can be consistently integrated once under the approximation $h\approx s \, \sin \theta$, which then leads to the model by \citet{han2014theoretical}.  

To conclude, we have established a quantitative model for contact line motion with normal stress, for arbitrary angles. The model provides a framework to test theoretical predictions and to interpret experimental observations for wetting flows of viscoelastic fluids. It consolidates the fundamental difference between the influence of normal stress on advancing and receding contact line motion, as previously reported in experimental studies. It would be of interest to further study the intricate transition from weak to strong viscoelasticity for advancing contact lines. In particular, one would like to see to what extent our findings are specific to the second-order fluid, and whether they can be observed in experiments or simulations based on more elaborate constitutive relations.

\bmhead{Acknowledgements}

The authors thank Jens Eggers for the fruitful discussions and the feedback on the draft of this manuscript. We thank Sander Huisman for the help with the initial numerical analysis in Mathematica. We also thank Uwe Thiele and Christopher Henkel for introducing us to the AUTO-07P code.

\section*{Statements and Declarations}

\bmhead{Funding} This work was supported by NWO through VICI grant no. 680-47-632. 

\bmhead{Declaration of competing interests} The authors declare that they have no known competing financial interests or personal relationships that could have appeared to influence the work reported in this paper.

\bmhead{Data availability} The authors declare that the data supporting the findings of this work are available within the paper. Numerical codes to generate the data may be requested from the authors.

\bmhead{Author ORCIDs}
Minkush Kansal: https://orcid.org/0000-0003-1584-3775;
\\
Charu Datt: https://orcid.org/0000-0002-9686-1774;
\\
Vincent Bertin: https://orcid.org/0000-0002-3139-8846;
\\
Jacco H. Snoeijer: https://orcid.org/0000-0001-6842-3024.

\begin{appendices}

\section{}
\label{sec:appendix}
In this Appendix, we provide an expression of the the second-order fluid model stress tensor with the corner flow velocity field. We make use of the fact that the corner flow velocity field is independent of $r$, such that the velocity gradient and shear rate tensors reduce to
\begin{equation}
	\nabla \pmb{u}= \begin{bmatrix}
		0 & 0 \\
		\dfrac{1}{r} \left(\dfrac{\partial u_r}{\partial \phi} - u_\phi \right) & \dfrac{1}{r} \left(\dfrac{\partial u_\phi}{\partial \phi} + u_r \right) 
	\end{bmatrix},
\end{equation}
\begin{equation}
	\label{eqn: gamma_tensor_planar_polar_0_r_derivatives}
	\dot{\pmb{\gamma}}= \begin{bmatrix}
		0  &  \dfrac{1}{r} \left(\dfrac{\partial u_r}{\partial \phi} - u_\phi \right) \\
		\dfrac{1}{r} \left(\dfrac{\partial u_r}{\partial \phi} - u_\phi \right) & \dfrac{2}{r} \left(\dfrac{\partial u_\phi}{\partial \phi} + u_r \right) 
	\end{bmatrix}. 
\end{equation}
In the steady state limit, the upper convective derivative term in \eqref{eqn:upper_convected_term} involves a linear combination of the three tensorial products $\pmb{u} \cdot \nabla \dot{\pmb{\gamma}}$, $\left(\nabla \pmb{u} \right)^T \cdot \dot{\pmb{\gamma}}$ and $\dot{\pmb{\gamma}} \cdot \nabla \pmb{u}$. Expressing these products in polar coordinates, we find
\begin{multline}
	\label{eqn:b2_term_part_1}
	\left(\pmb{u} \cdot \nabla \right) \dot{\pmb{\gamma}} =\\ \begin{bmatrix}
		-2 \dfrac{u_\phi}{r^2} \left(\dfrac{\partial u_r}{\partial \phi} - u_\phi\right) &  -\dfrac{u_r}{r^2} \left(\dfrac{\partial u_r}{\partial \phi} + u_\phi \right) + \dfrac{u_\phi}{r^2} \left(\dfrac{\partial^2 u_r}{\partial \phi^2} - 3 \dfrac{\partial u_\phi}{\partial \phi} \right) \\
		-\dfrac{u_r}{r^2} \left(\dfrac{\partial u_r}{\partial \phi} + u_\phi \right) + \dfrac{u_\phi}{r^2} \left(\dfrac{\partial^2 u_r}{\partial \phi^2} - 3 \dfrac{\partial u_\phi}{\partial \phi} \right) & \begin{split}
			2 \dfrac{u_\phi}{r^2} \left(2 \dfrac{\partial u_r}{\partial \phi} - u_\phi \right) - 2 \dfrac{u_r}{r^2} \left(\dfrac{\partial u_\phi}{\partial \phi} + u_r \right) \\ + 2 \dfrac{u_\phi}{r^2} \left( \dfrac{\partial^2 u_\phi}{\partial \phi^2}\right)
		\end{split}
	\end{bmatrix}, 
\end{multline}
\begin{equation}
	\label{eqn:b2_term_part_2}
	\left(\nabla \pmb{u} \right)^T \cdot \dot{\pmb{\gamma}}= \begin{bmatrix}
		\left(\dfrac{1}{r} \left(\dfrac{\partial u_r}{\partial \phi} - u_\phi\right) \right)^2  &  \dfrac{2}{r^2} \left(\dfrac{\partial u_\phi}{\partial \phi} + u_r \right) \left(\dfrac{\partial u_r}{\partial \phi} - u_\phi\right) \\
		\dfrac{1}{r^2} \left(\dfrac{\partial u_\phi}{\partial \phi} + u_r \right) \left(\dfrac{\partial u_r}{\partial \phi} - u_\phi\right) &  \dfrac{2}{r^2} \left(\dfrac{\partial u_\phi}{\partial \phi} + u_r \right)^2 
	\end{bmatrix},
\end{equation}
and 
\begin{equation}
	\label{eqn:b2_term_part_3}
	\dot{\pmb{\gamma}} \cdot \left(\nabla \pmb{u} \right)= \begin{bmatrix}
		\left(\dfrac{1}{r} \left(\dfrac{\partial u_r}{\partial \phi} - u_\phi\right) \right)^2 &  \dfrac{1}{r^2} \left(\dfrac{\partial u_\phi}{\partial \phi} + u_r \right) \left(\dfrac{\partial u_r}{\partial \phi} - u_\phi\right) \\
		\dfrac{2}{r^2} \left(\dfrac{\partial u_\phi}{\partial \phi} + u_r \right) \left(\dfrac{\partial u_r}{\partial \phi} - u_\phi\right) &  \dfrac{2}{r^2} \left(\dfrac{\partial u_\phi}{\partial \phi} + u_r \right)^2 
	\end{bmatrix},
\end{equation}
such that the upper convective derivative is given by
\begin{multline}
	\buildrel \nabla \over {\dot{\pmb{\gamma}}} =\\ 
	\begin{bmatrix}
		-2 \dfrac{u_\phi}{r^2} \left(\dfrac{\partial u_r}{\partial \phi} - u_\phi\right)- 2 \left(\dfrac{1}{r} \left(\dfrac{\partial u_r}{\partial \phi} - u_\phi\right) \right)^2 
		&  \begin{split}
			-\dfrac{u_r}{r^2} \left(\dfrac{\partial u_r}{\partial \phi} + u_\phi \right) + \dfrac{u_\phi}{r^2} \left(\dfrac{\partial^2 u_r}{\partial \phi^2} - 3 \dfrac{\partial u_\phi}{\partial \phi} \right) \\- \dfrac{3}{r^2} \left(\dfrac{\partial u_\phi}{\partial \phi} + u_r \right) \left(\dfrac{\partial u_r}{\partial \phi} - u_\phi\right) 
		\end{split}\\
		\begin{split}
			-\dfrac{u_r}{r^2} \left(\dfrac{\partial u_r}{\partial \phi} + u_\phi \right) + \dfrac{u_\phi}{r^2} \left(\dfrac{\partial^2 u_r}{\partial \phi^2} - 3 \dfrac{\partial u_\phi}{\partial \phi} \right) \\- \dfrac{3}{r^2} \left(\dfrac{\partial u_\phi}{\partial \phi} + u_r \right) \left(\dfrac{\partial u_r}{\partial \phi} - u_\phi\right) 
		\end{split} 
		& \begin{split}
			2 \dfrac{u_\phi}{r^2} \left(2 \dfrac{\partial u_r}{\partial \phi} - u_\phi \right) - 2 \dfrac{u_r}{r^2} \left(\dfrac{\partial u_\phi}{\partial \phi} + u_r \right) \\ + 2 \dfrac{u_\phi}{r^2} \left( \dfrac{\partial^2 u_\phi}{\partial \phi^2}\right)- \dfrac{4}{r^2} \left(\dfrac{\partial u_\phi}{\partial \phi} + u_r \right)^2 
		\end{split}
	\end{bmatrix}. \label{eqn:b2_term_part_expanded_matrix_form}
\end{multline}
Lastly, the quadratic term $\dot{\pmb{\gamma}} \cdot \dot{\pmb{\gamma}} $ in \eqref{eqn:tau_second_order_fluid} follows
\begin{equation}
	\label{eqn:gamma_dot_gamma_tensor}
	\dot{\pmb{\gamma}} \cdot \dot{\pmb{\gamma}} = \begin{bmatrix}
		\left( \dfrac{1}{r} \left(\dfrac{\partial u_r}{\partial \phi} - u_\phi \right) \right)^2 &  \dfrac{2}{r^2} \left(\dfrac{\partial u_r}{\partial \phi} - u_\phi \right) \left( \dfrac{\partial u_\phi}{\partial \phi} + u_r \right) \\
		\dfrac{2}{r^2} \left(\dfrac{\partial u_r}{\partial \phi} - u_\phi \right) \left( \dfrac{\partial u_\phi}{\partial \phi} + u_r \right) & \left( \dfrac{1}{r} \left(\dfrac{\partial u_r}{\partial \phi} - u_\phi \right) \right)^2 + \left( \dfrac{2}{r} \left(\dfrac{\partial u_\phi}{\partial \phi} + u_r \right) \right)^2 
	\end{bmatrix}.
\end{equation}
Expressing the shear component of the second-order fluid stress tensor $\left( \eta \dot{\pmb{\gamma}} -\frac{\psi_1}{2}\buildrel \nabla \over {\dot{\pmb{\gamma}}} + \psi_2 \dot{\pmb{\gamma}} \cdot \dot{\pmb{\gamma}}\right)_{r\phi}$ leads to Eq.~\eqref{eq:shear-stress_second-order}. 

\end{appendices}


\begin{thebibliography}{44}
\ifx \bisbn   \undefined \def \bisbn  #1{ISBN #1}\fi
\ifx \binits  \undefined \def \binits#1{#1}\fi
\ifx \bauthor  \undefined \def \bauthor#1{#1}\fi
\ifx \batitle  \undefined \def \batitle#1{#1}\fi
\ifx \bjtitle  \undefined \def \bjtitle#1{#1}\fi
\ifx \bvolume  \undefined \def \bvolume#1{\textbf{#1}}\fi
\ifx \byear  \undefined \def \byear#1{#1}\fi
\ifx \bissue  \undefined \def \bissue#1{#1}\fi
\ifx \bfpage  \undefined \def \bfpage#1{#1}\fi
\ifx \blpage  \undefined \def \blpage #1{#1}\fi
\ifx \burl  \undefined \def \burl#1{\textsf{#1}}\fi
\ifx \doiurl  \undefined \def \doiurl#1{\url{https://doi.org/#1}}\fi
\ifx \betal  \undefined \def \betal{\textit{et al.}}\fi
\ifx \binstitute  \undefined \def \binstitute#1{#1}\fi
\ifx \binstitutionaled  \undefined \def \binstitutionaled#1{#1}\fi
\ifx \bctitle  \undefined \def \bctitle#1{#1}\fi
\ifx \beditor  \undefined \def \beditor#1{#1}\fi
\ifx \bpublisher  \undefined \def \bpublisher#1{#1}\fi
\ifx \bbtitle  \undefined \def \bbtitle#1{#1}\fi
\ifx \bedition  \undefined \def \bedition#1{#1}\fi
\ifx \bseriesno  \undefined \def \bseriesno#1{#1}\fi
\ifx \blocation  \undefined \def \blocation#1{#1}\fi
\ifx \bsertitle  \undefined \def \bsertitle#1{#1}\fi
\ifx \bsnm \undefined \def \bsnm#1{#1}\fi
\ifx \bsuffix \undefined \def \bsuffix#1{#1}\fi
\ifx \bparticle \undefined \def \bparticle#1{#1}\fi
\ifx \barticle \undefined \def \barticle#1{#1}\fi
\bibcommenthead
\ifx \bconfdate \undefined \def \bconfdate #1{#1}\fi
\ifx \botherref \undefined \def \botherref #1{#1}\fi
\ifx \url \undefined \def \url#1{\textsf{#1}}\fi
\ifx \bchapter \undefined \def \bchapter#1{#1}\fi
\ifx \bbook \undefined \def \bbook#1{#1}\fi
\ifx \bcomment \undefined \def \bcomment#1{#1}\fi
\ifx \oauthor \undefined \def \oauthor#1{#1}\fi
\ifx \citeauthoryear \undefined \def \citeauthoryear#1{#1}\fi
\ifx \endbibitem  \undefined \def \endbibitem {}\fi
\ifx \bconflocation  \undefined \def \bconflocation#1{#1}\fi
\ifx \arxivurl  \undefined \def \arxivurl#1{\textsf{#1}}\fi
\csname PreBibitemsHook\endcsname

\bibitem[\protect\citeauthoryear{Scriven}{1988}]{scriven1988physics}
\begin{barticle}
\bauthor{\bsnm{Scriven}, \binits{L.}}:
\batitle{Physics and applications of dip coating and spin coating}.
\bjtitle{MRS Online Proceedings Library}
\bvolume{121}(\bissue{1}),
\bfpage{717}--\blpage{729}
(\byear{1988})
\end{barticle}
\endbibitem

\bibitem[\protect\citeauthoryear{Schyrr et~al.}{2014}]{schyrr2014biosensors}
\begin{barticle}
\bauthor{\bsnm{Schyrr}, \binits{B.}},
\bauthor{\bsnm{Pasche}, \binits{S.}},
\bauthor{\bsnm{Voirin}, \binits{G.}},
\bauthor{\bsnm{Weder}, \binits{C.}},
\bauthor{\bsnm{Simon}, \binits{Y.C.}},
\bauthor{\bsnm{Foster}, \binits{E.J.}}:
\batitle{Biosensors based on porous cellulose nanocrystal--poly (vinyl alcohol)
  scaffolds}.
\bjtitle{ACS applied materials \& interfaces}
\bvolume{6}(\bissue{15}),
\bfpage{12674}--\blpage{12683}
(\byear{2014})
\end{barticle}
\endbibitem

\bibitem[\protect\citeauthoryear{Riau et~al.}{2016}]{riau2016functionalization}
\begin{barticle}
\bauthor{\bsnm{Riau}, \binits{A.K.}},
\bauthor{\bsnm{Mondal}, \binits{D.}},
\bauthor{\bsnm{Setiawan}, \binits{M.}},
\bauthor{\bsnm{Palaniappan}, \binits{A.}},
\bauthor{\bsnm{Yam}, \binits{G.H.}},
\bauthor{\bsnm{Liedberg}, \binits{B.}},
\bauthor{\bsnm{Venkatraman}, \binits{S.S.}},
\bauthor{\bsnm{Mehta}, \binits{J.S.}}:
\batitle{Functionalization of the polymeric surface with bioceramic
  nanoparticles via a novel, nonthermal dip coating method}.
\bjtitle{ACS applied materials \& interfaces}
\bvolume{8}(\bissue{51}),
\bfpage{35565}--\blpage{35577}
(\byear{2016})
\end{barticle}
\endbibitem

\bibitem[\protect\citeauthoryear{Levich and Landau}{1942}]{levich1942dragging}
\begin{barticle}
\bauthor{\bsnm{Levich}, \binits{B.}},
\bauthor{\bsnm{Landau}, \binits{L.}}:
\batitle{Dragging of a liquid by a moving plate}.
\bjtitle{Acta Physicochim. URSS}
\bvolume{17},
\bfpage{141}
(\byear{1942})
\end{barticle}
\endbibitem

\bibitem[\protect\citeauthoryear{Qu{\'e}r{\'e}}{1999}]{quere1999fluid}
\begin{barticle}
\bauthor{\bsnm{Qu{\'e}r{\'e}}, \binits{D.}}:
\batitle{Fluid coating on a fiber}.
\bjtitle{Annual review of fluid mechanics}
\bvolume{31}(\bissue{1}),
\bfpage{347}--\blpage{384}
(\byear{1999})
\end{barticle}
\endbibitem

\bibitem[\protect\citeauthoryear{Rio and Boulogne}{2017}]{rio2017withdrawing}
\begin{barticle}
\bauthor{\bsnm{Rio}, \binits{E.}},
\bauthor{\bsnm{Boulogne}, \binits{F.}}:
\batitle{Withdrawing a solid from a bath: How much liquid is coated?}
\bjtitle{Advances in colloid and interface science}
\bvolume{247},
\bfpage{100}--\blpage{114}
(\byear{2017})
\end{barticle}
\endbibitem

\bibitem[\protect\citeauthoryear{Bertin et~al.}{2022}]{bertin2022enhanced}
\begin{barticle}
\bauthor{\bsnm{Bertin}, \binits{V.}},
\bauthor{\bsnm{Snoeijer}, \binits{J.H.}},
\bauthor{\bsnm{Rapha{\"e}l}, \binits{E.}},
\bauthor{\bsnm{Salez}, \binits{T.}}:
\batitle{Enhanced dip coating on a soft substrate}.
\bjtitle{Physical Review Fluids}
\bvolume{7}(\bissue{10}),
\bfpage{102002}
(\byear{2022})
\end{barticle}
\endbibitem

\bibitem[\protect\citeauthoryear{de~Gennes
  et~al.}{2004}]{gennes2004capillarity}
\begin{bbook}
\bauthor{\bsnm{Gennes}, \binits{P.-G.}},
\bauthor{\bsnm{Brochard-Wyart}, \binits{F.}},
\bauthor{\bsnm{Qu{\'e}r{\'e}}, \binits{D.}}:
\bbtitle{Capillarity and Wetting Phenomena: Drops, Bubbles, Pearls, Waves}.
\bpublisher{Springer},
\blocation{New York}
(\byear{2004})
\end{bbook}
\endbibitem

\bibitem[\protect\citeauthoryear{Eggers}{2005}]{eggers2005existence}
\begin{barticle}
\bauthor{\bsnm{Eggers}, \binits{J.}}:
\batitle{Existence of receding and advancing contact lines}.
\bjtitle{Physics of Fluids}
\bvolume{17}(\bissue{8}),
\bfpage{082106}
(\byear{2005})
\end{barticle}
\endbibitem

\bibitem[\protect\citeauthoryear{Snoeijer
  et~al.}{2007}]{snoeijer2007relaxation}
\begin{barticle}
\bauthor{\bsnm{Snoeijer}, \binits{J.H.}},
\bauthor{\bsnm{Andreotti}, \binits{B.}},
\bauthor{\bsnm{Delon}, \binits{G.}},
\bauthor{\bsnm{Fermigier}, \binits{M.}}:
\batitle{Relaxation of a dewetting contact line. part 1. a full-scale
  hydrodynamic calculation}.
\bjtitle{Journal of Fluid Mechanics}
\bvolume{579},
\bfpage{63}--\blpage{83}
(\byear{2007})
\end{barticle}
\endbibitem

\bibitem[\protect\citeauthoryear{Chan et~al.}{2012}]{chan2012theory}
\begin{botherref}
\oauthor{\bsnm{Chan}, \binits{T.S.}},
\oauthor{\bsnm{Snoeijer}, \binits{J.H.}},
\oauthor{\bsnm{Eggers}, \binits{J.}}:
Theory of the forced wetting transition.
Physics of fluids
\textbf{24}(7)
(2012)
\end{botherref}
\endbibitem

\bibitem[\protect\citeauthoryear{Galvagno
  et~al.}{2014}]{galvagno2014continuous}
\begin{barticle}
\bauthor{\bsnm{Galvagno}, \binits{M.}},
\bauthor{\bsnm{Tseluiko}, \binits{D.}},
\bauthor{\bsnm{Lopez}, \binits{H.}},
\bauthor{\bsnm{Thiele}, \binits{U.}}:
\batitle{Continuous and discontinuous dynamic unbinding transitions in drawn
  film flow}.
\bjtitle{Physical review letters}
\bvolume{112}(\bissue{13}),
\bfpage{137803}
(\byear{2014})
\end{barticle}
\endbibitem

\bibitem[\protect\citeauthoryear{Gupta et~al.}{2023}]{gupta2023experimental}
\begin{botherref}
\oauthor{\bsnm{Gupta}, \binits{C.}},
\oauthor{\bsnm{Choudhury}, \binits{A.}},
\oauthor{\bsnm{Chandrala}, \binits{L.D.}},
\oauthor{\bsnm{Dixit}, \binits{H.N.}}:
An experimental study of flow near an advancing contact line: a rigorous test
  of theoretical models.
arXiv preprint arXiv:2311.09560
(2023)
\end{botherref}
\endbibitem

\bibitem[\protect\citeauthoryear{De~Gennes}{1985}]{de1985wetting}
\begin{barticle}
\bauthor{\bsnm{De~Gennes}, \binits{P.-G.}}:
\batitle{Wetting: statics and dynamics}.
\bjtitle{Reviews of modern physics}
\bvolume{57}(\bissue{3}),
\bfpage{827}
(\byear{1985})
\end{barticle}
\endbibitem

\bibitem[\protect\citeauthoryear{Bonn et~al.}{2009}]{bonn2009wetting}
\begin{barticle}
\bauthor{\bsnm{Bonn}, \binits{D.}},
\bauthor{\bsnm{Eggers}, \binits{J.}},
\bauthor{\bsnm{Indekeu}, \binits{J.}},
\bauthor{\bsnm{Meunier}, \binits{J.}},
\bauthor{\bsnm{Rolley}, \binits{E.}}:
\batitle{Wetting and spreading}.
\bjtitle{Reviews of modern physics}
\bvolume{81}(\bissue{2}),
\bfpage{739}
(\byear{2009})
\end{barticle}
\endbibitem

\bibitem[\protect\citeauthoryear{Snoeijer and
  Andreotti}{2013}]{snoeijer2013moving}
\begin{barticle}
\bauthor{\bsnm{Snoeijer}, \binits{J.H.}},
\bauthor{\bsnm{Andreotti}, \binits{B.}}:
\batitle{Moving contact lines: scales, regimes, and dynamical transitions}.
\bjtitle{Annual review of fluid mechanics}
\bvolume{45},
\bfpage{269}--\blpage{292}
(\byear{2013})
\end{barticle}
\endbibitem

\bibitem[\protect\citeauthoryear{Bird et~al.}{1987}]{bird1987dynamics}
\begin{botherref}
\oauthor{\bsnm{Bird}, \binits{R.B.}},
\oauthor{\bsnm{Armstrong}, \binits{R.C.}},
\oauthor{\bsnm{Hassager}, \binits{O.}}:
Dynamics of polymeric liquids. vol. 1: Fluid mechanics
(1987)
\end{botherref}
\endbibitem

\bibitem[\protect\citeauthoryear{Tanner}{2000}]{tanner2000engineering}
\begin{bbook}
\bauthor{\bsnm{Tanner}, \binits{R.I.}}:
\bbtitle{Engineering Rheology}
vol. \bseriesno{52}.
\bpublisher{Oxford University Press Inc.},
\blocation{New York}
(\byear{2000})
\end{bbook}
\endbibitem

\bibitem[\protect\citeauthoryear{Bartolo et~al.}{2007}]{bartolo2007dynamics}
\begin{barticle}
\bauthor{\bsnm{Bartolo}, \binits{D.}},
\bauthor{\bsnm{Boudaoud}, \binits{A.}},
\bauthor{\bsnm{Narcy}, \binits{G.}},
\bauthor{\bsnm{Bonn}, \binits{D.}}:
\batitle{Dynamics of non-newtonian droplets}.
\bjtitle{Physical review letters}
\bvolume{99}(\bissue{17}),
\bfpage{174502}
(\byear{2007})
\end{barticle}
\endbibitem

\bibitem[\protect\citeauthoryear{Datt et~al.}{2022}]{datt2022thin}
\begin{barticle}
\bauthor{\bsnm{Datt}, \binits{C.}},
\bauthor{\bsnm{Kansal}, \binits{M.}},
\bauthor{\bsnm{Snoeijer}, \binits{J.H.}}:
\batitle{A thin-film equation for a viscoelastic fluid, and its application to
  the landau--levich problem}.
\bjtitle{Journal of Non-Newtonian Fluid Mechanics}
\bvolume{305},
\bfpage{104816}
(\byear{2022})
\end{barticle}
\endbibitem

\bibitem[\protect\citeauthoryear{Kansal et~al.}{2024}]{kansal2024viscoelastic}
\begin{barticle}
\bauthor{\bsnm{Kansal}, \binits{M.}},
\bauthor{\bsnm{Bertin}, \binits{V.}},
\bauthor{\bsnm{Datt}, \binits{C.}},
\bauthor{\bsnm{Eggers}, \binits{J.}},
\bauthor{\bsnm{Snoeijer}, \binits{J.H.}}:
\batitle{Viscoelastic wetting: Cox--voinov theory with normal stress effects}.
\bjtitle{Journal of Fluid Mechanics}
\bvolume{985},
\bfpage{17}
(\byear{2024})
\end{barticle}
\endbibitem

\bibitem[\protect\citeauthoryear{Seevaratnam
  et~al.}{2007}]{seevaratnam2007dynamic}
\begin{botherref}
\oauthor{\bsnm{Seevaratnam}, \binits{G.}},
\oauthor{\bsnm{Suo}, \binits{Y.}},
\oauthor{\bsnm{Ram{\'e}}, \binits{E.}},
\oauthor{\bsnm{Walker}, \binits{L.}},
\oauthor{\bsnm{Garoff}, \binits{S.}}:
Dynamic wetting of shear thinning fluids.
Physics of Fluids
\textbf{19}(1)
(2007)
\end{botherref}
\endbibitem

\bibitem[\protect\citeauthoryear{Wei et~al.}{2009}]{wei2009dynamic}
\begin{barticle}
\bauthor{\bsnm{Wei}, \binits{Y.}},
\bauthor{\bsnm{Rame}, \binits{E.}},
\bauthor{\bsnm{Walker}, \binits{L.}},
\bauthor{\bsnm{Garoff}, \binits{S.}}:
\batitle{Dynamic wetting with viscous newtonian and non-newtonian fluids}.
\bjtitle{Journal of Physics: Condensed Matter}
\bvolume{21}(\bissue{46}),
\bfpage{464126}
(\byear{2009})
\end{barticle}
\endbibitem

\bibitem[\protect\citeauthoryear{Smith and Sharp}{2014}]{smith2014origin}
\begin{barticle}
\bauthor{\bsnm{Smith}, \binits{M.}},
\bauthor{\bsnm{Sharp}, \binits{J.}}:
\batitle{Origin of contact line forces during the retraction of dilute polymer
  solution drops}.
\bjtitle{Langmuir}
\bvolume{30}(\bissue{19}),
\bfpage{5455}--\blpage{5459}
(\byear{2014})
\end{barticle}
\endbibitem

\bibitem[\protect\citeauthoryear{Shin and Kim}{2015}]{shin2015contact}
\begin{barticle}
\bauthor{\bsnm{Shin}, \binits{H.}},
\bauthor{\bsnm{Kim}, \binits{C.}}:
\batitle{Contact line motion of polymer solution inside capillary}.
\bjtitle{Journal of Non-Newtonian Fluid Mechanics}
\bvolume{218},
\bfpage{62}--\blpage{70}
(\byear{2015})
\end{barticle}
\endbibitem

\bibitem[\protect\citeauthoryear{Yue and Feng}{2012}]{yue2012phase}
\begin{barticle}
\bauthor{\bsnm{Yue}, \binits{P.}},
\bauthor{\bsnm{Feng}, \binits{J.J.}}:
\batitle{Phase-field simulations of dynamic wetting of viscoelastic fluids}.
\bjtitle{Journal of Non-Newtonian Fluid Mechanics}
\bvolume{189},
\bfpage{8}--\blpage{13}
(\byear{2012})
\end{barticle}
\endbibitem

\bibitem[\protect\citeauthoryear{Wang et~al.}{2015}]{wang2015dynamic}
\begin{barticle}
\bauthor{\bsnm{Wang}, \binits{Y.}},
\bauthor{\bsnm{Minh}, \binits{D.-Q.}},
\bauthor{\bsnm{Amberg}, \binits{G.}}:
\batitle{Dynamic wetting of viscoelastic droplets}.
\bjtitle{Physical Review E}
\bvolume{92}(\bissue{4}),
\bfpage{043002}
(\byear{2015})
\end{barticle}
\endbibitem

\bibitem[\protect\citeauthoryear{Morozov and
  Spagnolie}{2015}]{morozov2015introduction}
\begin{botherref}
\oauthor{\bsnm{Morozov}, \binits{A.}},
\oauthor{\bsnm{Spagnolie}, \binits{S.E.}}:
Introduction to complex fluids.
Complex Fluids in Biological Systems: Experiment, Theory, and Computation,
3--52
(2015)
\end{botherref}
\endbibitem

\bibitem[\protect\citeauthoryear{De~Corato et~al.}{2016}]{decorato16b}
\begin{barticle}
\bauthor{\bsnm{De~Corato}, \binits{M.}},
\bauthor{\bsnm{Greco}, \binits{F.}},
\bauthor{\bsnm{Maffettone}, \binits{P.L.}}:
\batitle{Reply to ``{C}omment on `{L}ocomotion of a microorganism in weakly
  viscoelastic liquids' ''}.
\bjtitle{Phys. Rev. E}
\bvolume{94},
\bfpage{057102}
(\byear{2016})
\doiurl{10.1103/PhysRevE.94.057102}
\end{barticle}
\endbibitem

\bibitem[\protect\citeauthoryear{Snoeijer}{2006}]{snoeijer2006free}
\begin{botherref}
\oauthor{\bsnm{Snoeijer}, \binits{J.H.}}:
Free-surface flows with large slopes: Beyond lubrication theory.
Physics of Fluids
\textbf{18}(2)
(2006)
\end{botherref}
\endbibitem

\bibitem[\protect\citeauthoryear{Huh and Scriven}{1971}]{huh1971hydrodynamic}
\begin{barticle}
\bauthor{\bsnm{Huh}, \binits{C.}},
\bauthor{\bsnm{Scriven}, \binits{L.E.}}:
\batitle{Hydrodynamic model of steady movement of a solid/liquid/fluid contact
  line}.
\bjtitle{Journal of colloid and interface science}
\bvolume{35}(\bissue{1}),
\bfpage{85}--\blpage{101}
(\byear{1971})
\end{barticle}
\endbibitem

\bibitem[\protect\citeauthoryear{Cox}{1986}]{cox_1986}
\begin{barticle}
\bauthor{\bsnm{Cox}, \binits{R.G.}}:
\batitle{The dynamics of the spreading of liquids on a solid surface. part 1.
  viscous flow}.
\bjtitle{Journal of Fluid Mechanics}
\bvolume{168},
\bfpage{169}--\blpage{194}
(\byear{1986})
\doiurl{10.1017/S0022112086000332}
\end{barticle}
\endbibitem

\bibitem[\protect\citeauthoryear{Han and Kim}{2014}]{han2014theoretical}
\begin{barticle}
\bauthor{\bsnm{Han}, \binits{J.}},
\bauthor{\bsnm{Kim}, \binits{C.}}:
\batitle{Theoretical and experimental studies on the contact line motion of
  second-order fluid}.
\bjtitle{Rheologica Acta}
\bvolume{53},
\bfpage{55}--\blpage{66}
(\byear{2014})
\doiurl{10.1007/s00397-013-0743-1}
\end{barticle}
\endbibitem

\bibitem[\protect\citeauthoryear{Tanner}{1966}]{tanner1966plane}
\begin{barticle}
\bauthor{\bsnm{Tanner}, \binits{R.}}:
\batitle{Plane creeping flows of incompressible second-order fluids}.
\bjtitle{Physics of Fluids}
\bvolume{9}(\bissue{6}),
\bfpage{1246}
(\byear{1966})
\end{barticle}
\endbibitem

\bibitem[\protect\citeauthoryear{Tanner}{1989}]{tanner1989some}
\begin{barticle}
\bauthor{\bsnm{Tanner}, \binits{R.}}:
\batitle{Some extended giesekus-type theorems for non-newtonian flows}.
\bjtitle{Rheologica acta}
\bvolume{28},
\bfpage{449}--\blpage{452}
(\byear{1989})
\end{barticle}
\endbibitem

\bibitem[\protect\citeauthoryear{Happel and Brenner}{1983}]{happel1983low}
\begin{bbook}
\bauthor{\bsnm{Happel}, \binits{J.}},
\bauthor{\bsnm{Brenner}, \binits{H.}}:
\bbtitle{Low Reynolds Number Hydrodynamics: with Special Applications to
  Particulate Media}
vol. \bseriesno{1}.
\bpublisher{Martinus Nijhoff Publishers},
\blocation{The Hague}
(\byear{1983})
\end{bbook}
\endbibitem

\bibitem[\protect\citeauthoryear{Chan et~al.}{2013}]{chan2013hydrodynamics}
\begin{botherref}
\oauthor{\bsnm{Chan}, \binits{T.S.}},
\oauthor{\bsnm{Srivastava}, \binits{S.}},
\oauthor{\bsnm{Marchand}, \binits{A.}},
\oauthor{\bsnm{Andreotti}, \binits{B.}},
\oauthor{\bsnm{Biferale}, \binits{L.}},
\oauthor{\bsnm{Toschi}, \binits{F.}},
\oauthor{\bsnm{Snoeijer}, \binits{J.H.}}:
Hydrodynamics of air entrainment by moving contact lines.
Physics of fluids
\textbf{25}(7)
(2013)
\end{botherref}
\endbibitem

\bibitem[\protect\citeauthoryear{Chan et~al.}{2020}]{chan2020cox}
\begin{barticle}
\bauthor{\bsnm{Chan}, \binits{T.S.}},
\bauthor{\bsnm{Kamal}, \binits{C.}},
\bauthor{\bsnm{Snoeijer}, \binits{J.H.}},
\bauthor{\bsnm{Sprittles}, \binits{J.E.}},
\bauthor{\bsnm{Eggers}, \binits{J.}}:
\batitle{Cox--voinov theory with slip}.
\bjtitle{Journal of fluid mechanics}
\bvolume{900},
\bfpage{8}
(\byear{2020})
\end{barticle}
\endbibitem

\bibitem[\protect\citeauthoryear{Doedel et~al.}{2007}]{doedel2007auto}
\begin{botherref}
\oauthor{\bsnm{Doedel}, \binits{E.J.}},
\oauthor{\bsnm{Champneys}, \binits{A.R.}},
\oauthor{\bsnm{Dercole}, \binits{F.}},
\oauthor{\bsnm{Fairgrieve}, \binits{T.F.}},
\oauthor{\bsnm{Kuznetsov}, \binits{Y.A.}},
\oauthor{\bsnm{Oldeman}, \binits{B.}},
\oauthor{\bsnm{Paffenroth}, \binits{R.}},
\oauthor{\bsnm{Sandstede}, \binits{B.}},
\oauthor{\bsnm{Wang}, \binits{X.}},
\oauthor{\bsnm{Zhang}, \binits{C.}}:
Auto-07p: Continuation and bifurcation software for ordinary differential
  equations
(2007)
\end{botherref}
\endbibitem

\bibitem[\protect\citeauthoryear{Guyard et~al.}{2021}]{guyard2021near}
\begin{barticle}
\bauthor{\bsnm{Guyard}, \binits{G.}},
\bauthor{\bsnm{Vilquin}, \binits{A.}},
\bauthor{\bsnm{Sanson}, \binits{N.}},
\bauthor{\bsnm{Jouenne}, \binits{S.}},
\bauthor{\bsnm{Restagno}, \binits{F.}},
\bauthor{\bsnm{McGraw}, \binits{J.D.}}:
\batitle{Near-surface rheology and hydrodynamic boundary condition of
  semi-dilute polymer solutions}.
\bjtitle{Soft Matter}
\bvolume{17}(\bissue{14}),
\bfpage{3765}--\blpage{3774}
(\byear{2021})
\end{barticle}
\endbibitem

\bibitem[\protect\citeauthoryear{Lorenceau et~al.}{2004}]{lorenceau2004air}
\begin{barticle}
\bauthor{\bsnm{Lorenceau}, \binits{{\'E}.}},
\bauthor{\bsnm{Qu{\'e}r{\'e}}, \binits{D.}},
\bauthor{\bsnm{Eggers}, \binits{J.}}:
\batitle{Air entrainment by a viscous jet plunging into a bath}.
\bjtitle{Physical review letters}
\bvolume{93}(\bissue{25}),
\bfpage{254501}
(\byear{2004})
\end{barticle}
\endbibitem

\bibitem[\protect\citeauthoryear{Marchand et~al.}{2012}]{marchand2012air}
\begin{barticle}
\bauthor{\bsnm{Marchand}, \binits{A.}},
\bauthor{\bsnm{Chan}, \binits{T.S.}},
\bauthor{\bsnm{Snoeijer}, \binits{J.H.}},
\bauthor{\bsnm{Andreotti}, \binits{B.}}:
\batitle{Air entrainment by contact lines of a solid plate plunged into a
  viscous fluid}.
\bjtitle{Physical review letters}
\bvolume{108}(\bissue{20}),
\bfpage{204501}
(\byear{2012})
\end{barticle}
\endbibitem

\bibitem[\protect\citeauthoryear{Blake and Ruschak}{1979}]{blake1979maximum}
\begin{barticle}
\bauthor{\bsnm{Blake}, \binits{T.D.}},
\bauthor{\bsnm{Ruschak}, \binits{K.J.}}:
\batitle{A maximum speed of wetting}.
\bjtitle{Nature}
\bvolume{282}(\bissue{5738}),
\bfpage{489}--\blpage{491}
(\byear{1979})
\end{barticle}
\endbibitem

\bibitem[\protect\citeauthoryear{Sedev and Petrov}{1991}]{sedev1991critical}
\begin{barticle}
\bauthor{\bsnm{Sedev}, \binits{R.}},
\bauthor{\bsnm{Petrov}, \binits{J.}}:
\batitle{The critical condition for transition from steady wetting to film
  entrainment}.
\bjtitle{Colloids and surfaces}
\bvolume{53}(\bissue{1}),
\bfpage{147}--\blpage{156}
(\byear{1991})
\end{barticle}
\endbibitem

\end{thebibliography}
\end{document}